\documentclass[envcountsame,runningheads]{llncs}
\usepackage{xspace}
\usepackage{amsmath}
\usepackage{amssymb}
\usepackage{graphicx}
\DeclareFontFamily{U}{mathb}{\hyphenchar\font45}
\DeclareFontShape{U}{mathb}{m}{n}{
      <5> <6> <7> <8> <9> <10> gen * mathb
      <10.95> mathb10 <12> <14.4> <17.28> <20.74> <24.88> mathb12
      }{}
\DeclareSymbolFont{mathb}{U}{mathb}{m}{n}
\DeclareMathSymbol{\precneq}{3}{mathb}{"AC}
\newcommand{\Nreduceq}{\reduceq_{\text{n}}}
\newcommand{\Nequiv}{\equiv_{\text{n}}}

\newcommand{\reduceq}{\preceq}
\newcommand{\reducneq}{\precneq}

\newcommand{\reduceqt}{\reduceq_\text{t}}
\newcommand{\reducneqt}{\precneq_\text{t}}
\newcommand{\notreduceqt}{\npreceq_\text{t}}
\newcommand{\BCSS}{BCSS\xspace} 
\newcommand{\IR}{\mathbb{R}}
\newcommand{\IRc}{\mathbb{R}_c}

\newcommand{\IQ}{\mathbb{Q}}

\newcommand{\IN}{\mathbb{N}}
\newcommand{\dom}{\operatorname{dom}}
\newcommand{\bin}{\operatorname{bin}}
\newcommand{\id}{\operatorname{id}}
\newcommand{\myjoin}{\wedge}  
\DeclareMathOperator*{\mysup}{\smash{\sup}}
\DeclareMathOperator*{\myinf}{\smash{\inf}}
\DeclareMathOperator*{\mytheta}{\Theta}
\DeclareMathOperator*{\ulim}{ulim}
\newcommand{\calO}{\mathcal{O}}
\newcommand{\calM}{\mathcal{M}}
\newcommand{\Heavi}{h}
\newcommand{\barH}{\overline{\Heavi}}
\newcommand{\RDelta}[1]{\Delta_{#1}\IR}
\newcommand{\RSigma}[1]{\Sigma_{#1}\IR}
\newcommand{\RPi}[1]{\Pi_{#1}\IR}
\newcommand{\person}[1]{\textsc{#1}}
\newcommand{\aname}[1]{\textsf{#1}}
\newcommand{\mycite}[2]{\cite[\textsc{#1}]{#2}}
\newcommand{\myto}{\!\to\!}
\newcommand{\myrho}{\rho}
\newcommand{\myl}{{\scriptscriptstyle<}}
\newcommand{\myg}{{\scriptscriptstyle>}}
\newcommand{\myrhol}{\myrho_{\raisebox{0.2ex}{$\myl$}}}
\newcommand{\myrhog}{\myrho_{\raisebox{0.2ex}{$\myg$}}}
\newcommand{\myrhon}{\myrho_{\text{Cn}}}

\newcommand{\myrhob}{\myrho_{\text{b},2}}
\newcommand{\myrhoh}{\myrho_{\text{H}}}
\newcommand{\tildemyrhol}{\tilde\myrho_{\raisebox{0.2ex}{$\myl$}}}
\newcommand{\hatmyrhol}{\hat\myrho_{\raisebox{0.2ex}{$\myl$}}}
\newcommand{\nuQ}{\nu_{\IQ}}
\newcommand{\Card}[1]{|{#1}|}
\newcommand{\Gdelta}{\text{G}_{\delta}}
\newcommand{\lcp}{\operatorname{lcp}}
\newcommand{\substr}{\sqsubseteq}
\newcommand{\supstr}{\sqsupseteq}
\newcommand{\LSC}{\operatorname{LSC}}
\newcommand{\MLSC}[1]{\operatorname{MLSC}(#1)}
\newcommand{\COMMENTED}[1]{}

\spnewtheorem{observation}[theorem]{Observation}{\bfseries}{\itshape}
\spnewtheorem{fact}[theorem]{Fact}{\bfseries}{\itshape}
\spnewtheorem{myclaim}[theorem]{Claim}{\bfseries}{\itshape}
\spnewtheorem{scholium}[theorem]{Scholium}{\bfseries}{\itshape}
\newtheorem{scholiumf}[theorem]{Scholium\footnotemark}
\begin{document}
\title{Real Hypercomputation and Continuity\thanks{An extended abstract
of this work, mostly lacking proofs, has appeared as \cite{CIE}.}}
\titlerunning{Real Hypercomputation and Continuity}
\author{Martin Ziegler\thanks{supported by the German Research Fundation
\textsf{DFG} with project \texttt{Zi1009/1-1}}}
\authorrunning{Martin Ziegler}
\institute{University of Paderborn, 33095 GERMANY}
\maketitle
\def\thefootnote{\fnsymbol{footnote}}
\addtocounter{footnote}{2}
\begin{abstract}
By the sometimes so-called \emph{Main Theorem} of
Recursive Analysis, every computable real function
is necessarily continuous.
We wonder whether and which kinds of \emph{hyper}computation
allow for the effective evaluation of also discontinuous $f:\IR\to\IR$.
More precisely the present work considers the following three
super-Turing notions of real function computability:
\begin{itemize}
\item relativized computation; \quad specifically
  given oracle access to the Halting Problem
  $\emptyset'$ or its jump $\emptyset''$;
\item encoding input $x\in\IR$ and/or output $y=f(x)$ in weaker
  ways also related to the Arithmetic Hierarchy;
\item non-deterministic computation.
\end{itemize}
It turns out that any $f:\IR\to\IR$ computable in the
first or second sense is still necessarily continuous
whereas the third type of hypercomputation
does provide the required power to evaluate for instance
the discontinuous Heaviside function.
\end{abstract}
\section{Motivation}
What does it mean for a Turing Machine,
capable of operating only on discrete objects,
to compute a real number $x$:

\smallskip
\noindent\begin{tabular}{rl}
$\pmb{\myrhob}$\textbf{:} & To determine its binary expansion, i.e.,
 to decide $A\subseteq\IN$ with $x=\sum\limits_{n\in A} 2^{-n}$ ?\\
$\pmb{\myrhon}$\textbf{:} & To compute a sequence $(q_n)$ of rational numbers
  eventually converging to $x$?\\[0.7ex]
$\pmb{\myrho}$\textbf{:} & To compute a \emph{fast} convergent sequence
  $(q_n)\subseteq\IQ$ for $x$, ~i.e. with $|x-q_n|$ \\
& $\leq2^{-n}$ ~ (in other words:
 to approximate $x$ with effective error bounds)?\\[0.5ex]
$\pmb{\myrhol}$\textbf{:} & To approximate $x$ from below, i.e., to
  compute $(q_n)$ such that $x=\sup_n q_n$ ?
\end{tabular}

\medskip
\noindent
All these notions make sense in being closed under arithmetic
operations like addition and multiplication. In fact they are
well (known equivalent to variants) studied in literature\footnote{%
Their above names by indexed Greek letters are taken from
\mycite{Section~4.1}{Weihrauch}.};
e.g. \cite{Turing}, \cite{Naive}, \cite{Turing2}, \cite{Weihrauch} in order.

\medskip
Now what does it mean for a Turing Machine $\calM$
to compute a real function $f:\IR\to\IR$?
Most naturally it means that $\calM$ realizes effective
evaluation $x\mapsto f(x)$ in that, upon input of $x\in\IR$
given in one of the above ways, it outputs $y=f(x)$ also
in one (not necessarily the same) of the above ways.
And, again, many possible combinations have already been investigated.
For instance the standard notion of real function computation
in Recursive Analysis \cite{Grzegorczyk,PER,Ko,Weihrauch} refers
(or is equivalent) to input and output given according to $\myrho$.
Here, the \aname{Main Theorem} of Computable Analysis implies that
any computable $f$ will necessarily be continuous
\mycite{Theorem~4.3.1}{Weihrauch}.

We are interested in ways of lifting this restriction,
that is, in the following
\begin{question}
Does hypercomputation in some sense permit
the computational evaluation of (at least certain)
discontinuous real functions?
\end{question}
That is related to the \aname{Church-Turing Hypothesis}:
A Turing Machine's ability to simulate \emph{every} physical
process would imply all such processes to behave continuously%
---a property \person{G.~Leibniz} was convinced of 
(``\emph{Natura non facit saltus}'')
but which we nowadays know to be violated for instance
by the \aname{Quantum Hall Effect} awarded a Nobel Prize in 1985.
Since this (nor any other) discontinuous physical process cannot
be simulated on a classical Turing Machine, it constitutes a putative
candidate for a system capable of realizing hypercomputation.

\subsection{Summary}
The standard (and indeed the most general) way of turning a
Turing Machine into a hypercomputer is to grant it access
to an oracle like, say, the
Halting Problem $\emptyset'$ or its iterated jumps like $\emptyset''$
and $\emptyset^{(d)}$
in \person{Kleene}'s \aname{Arithmetic Hierarchy}.
However regarding computational evaluation of real functions,
closer inspection in Section~\ref{secOracles}
reveals that this Main Theorem
relies solely on information rather than recursion theoretic arguments
and therefore requires continuity also
for oracle-computable real functions
with respect to input and output of form $\myrho$.
(For the special case of an $\emptyset'$--oracle,
this had been observed in \mycite{Theorem~16}{Ho}.)

\medskip
A second idea, applicable to real but not to discrete computability,
changes the input and output representation
for $x$ and $y=f(x)$ from $\myrho$ to a weaker
form like, say, $\myrhon$. This relates to
the Arithmetic Hierarchy, too, however in a 
different way: Computing $x$ in the sense
of $\myrhon$ is equivalent to computing
$x$ in the sense of $\myrho$ \mycite{Theorem~9}{Ho}
\emph{relative} (i.e., given access) to the
Halting Problem $\emptyset'$ and thus suggests 
to write $\myrho':=\myrhon$.
Most promisingly, the Main Theorem
\mycite{Corollary~3.2.12}{Weihrauch}
which requires continuity of $(\myrho\to\myrho)$--computable
real functions applies to $\myrho$ but not to
$\myrho'$ because the latter lacks
the technical property of \emph{admissibility}.

It therefore came to quite a surprise when
\person{Brattka} and \person{Hertling} established
that any $(\myrho'\to\myrho')$--computable $f$
(that is, with respect to input $x$
and output $f(x)$ encoded according to $\myrhon$) still
satisfies continuity;
see \cite[\textsc{Exercise~4.1.13}d]{Weihrauch}
or \mycite{Section~6}{Naive}.

Section~\ref{secWeak} contains an extension of this and
a series of related results.
Specifically we manage to prove that
continuity is necessary for
$(\myrho''\to\myrho'')$--computability of $f$;
here, ~$\myrho\:\reducneq\:
\myrho'\:\reducneq\:\myrho''\:\reducneq\ldots$~
denote the first levels of an entire hierarchy of real
number representations explained in Lemma~\ref{lemRepr}
which emerge naturally from
the Real Arithmetic Hierarchy of
\person{Weihrauch} and \person{Zheng} \cite{Zheng}.

\medskip
In Section~\ref{secHierarchies}, we closer investigate
the two approaches to real function hypercomputation.
Specifically it is established (Section~\ref{secWeakly})
that the hierarchy of real number representation actually
does yield a hierarchy of weakly computable real functions.
Furthermore a comparison of both oracle-supported and weakly
computable (and each hence necessarily continuous)
real functions in Section~\ref{secWeier}
reveals a relativized version of the
Effective Weierstra\ss{} Theorem to fail.

\medskip
Our third approach to real hypercomputation (Section~\ref{secNondet}) 
finally allows the Turing Machines
under consideration to behave \emph{non}deterministically.
Remarkably and in contrast to the classical (Type-1) theory,
this does significantly increase their principal capabilities.
For example, all quasi-strongly $\delta$--$\IQ$--analytic functions
in the sense of \person{Chadzelek} and \person{Hotz} \cite{Hotz}%
---and in particular many discontinuous real functions---now become
computable as well as conversion among the
aforementioned representations $\myrhon$ and $\myrhob$.
\section{Arithmetic Hierarchy and Reals} \label{secRAH}
In \cite{Ho}, \person{Ho} observed an
interesting relation between computability of a real number $x$
in the respective senses of $\myrho$ and $\myrhon$ in terms of oracles:
$x=\lim_n q_n$ for an (eventually convergent)
computable rational sequence $(q_n)$ ~iff~ $x$
admits a \emph{fast} convergent rational sequence
computable with oracle $\emptyset'$,
that is, a sequence $(p_m)\subseteq\IQ$
recursive in $\emptyset'$ with $|x-p_m|\leq2^{-m}$.
This suggests to use $\myrho'$ synonymously for $\myrhon$; 
and denoting by $\RDelta{1}=\IRc$ the set of
reals computable in the sense of Recursive Analysis
(that is with respect to $\myrho$), it is therefore natural
to write, in analogy to
\person{Kleene}'s classical \aname{Arithmetic Hierarchy},
$\RDelta{2}$ for the set of all $x\in\IR$ computable
with respect to $\myrho'$.
\person{Weihrauch} and \person{Zheng} extended
these considerations and obtained for instance
\mycite{Corollary~7.3}{Zheng} the following characterization
of $\RDelta{3}$: A real $x\in\IR$
admits a 
fast convergent rational sequence
recursive in $\emptyset''$ iff $x$ is computable in the
sense of $\myrho''$ defined as follows:

\smallskip
\noindent\begin{tabular}{rl}
\hspace*{1.5em}$\pmb{\myrho''}$\textbf{:} & ~ $\displaystyle
  x=\lim\nolimits_i \lim\nolimits_j q_{\scriptscriptstyle\langle i,j\rangle}$
  ~for some computable rational sequence $(q_n)$
\end{tabular}

\smallskip\noindent
where $\langle\cdots\rangle:\IN^*\to\IN$ 
denotes some fixed computable pairing or, more generally,
tupling function.
Similarly, $\RSigma{1}$ contains of all $x\in\IR$
computable with respect to $\myrhol$
whereas $\RSigma{2}$ includes all $x$ computable
in the sense of $\myrhol'$ defined as follows:

\smallskip
\noindent\begin{tabular}{rl}
\hspace*{1.5em}$\pmb{\myrhol'}$\textbf{:} & ~
$\displaystyle x=\mysup\nolimits_i \myinf\nolimits_j
q_{\scriptscriptstyle\langle i,j\rangle}$
~for some computable rational sequence $(q_n)$.
\end{tabular}

\smallskip\noindent
To $\RSigma{2}$ belongs for instance the radius of 
convergence $r=1/\limsup_{n\to\infty}\sqrt[n]{a_n}$ of
a computable power series $\sum_{n=0}^\infty a_nx^n$
\mycite{Theorem~6.2}{Zheng}. More generally we
take from \cite[\textsc{Definition~7.1} and 
\textsc{Corollary~7.3}]{Zheng} the following
\begin{definition}[Real Arithmetic Hierarchy]
\label{defRAH} Let $d=0,1,2,\ldots$
\\\begin{tabular}{rl}
$\pmb{\myrhol^{(d)}}$\textbf{:} &
  $\RSigma{d+1}$ consists of all $x\in\IR$ of the form
  ~ $\displaystyle x\;=\;
    \mysup_{n_1} \; \myinf_{n_2} \;\ldots \; \mytheta_{n_{d+1}}
      \;q_{\scriptscriptstyle\langle n_1,\ldots,n_{d+1}\rangle}$
 \\[-1.1ex] & for a computable rational sequence $(q_n)$,  
 \\ & where $\mytheta\!\!=\!\mysup$ or $\mytheta\!\!=\!\myinf$
 depending on $d$'s parity; \\
$\pmb{\myrhog^{(d)}}$\textbf{:} & $\RPi{d+1}$ similarly for
  ~$\displaystyle x=\myinf_{n_1} \mysup_{n_2} \;\ldots$ \\
$\pmb{\myrho^{(d)}}$\textbf{:} & $\RDelta{d+1}$ contains all $x\in\IR$
  of the form
  \quad$\displaystyle x\;=\;
    \lim_{n_1} \; \lim_{n_2} \;\ldots \; \lim_{n_d}
      \;q_{\scriptscriptstyle\langle n_1,\ldots,n_d\rangle}$
 \\[-1.1ex] & for a computable rational sequence $(q_n)$.
\end{tabular}
\end{definition}
(For an extension to levels beyond $\omega$ see \cite{Barmpalias}\ldots)

The close analogy between the discrete and this real variant of the
Arithmetic Hierarchy is expressed in \cite{Zheng}
by a variety of elegant results like, e.g.,
\begin{fact} \label{facRAH} 
\begin{enumerate}
\item[a)] $x\in\RDelta{d}$ ~iff~ deciding its binary expansion is in
  $\Delta_d$.
\item[b)] $x$ is computable with respect to $\myrho^{(d)}$ \\
  ~iff~ there is a
  $\emptyset^{(d)}$--computable fast convergent rational sequence
    for $x$.
\item[c)] $x$ is computable with respect to
  $\myrhol^{(d)}$ \\
  ~iff~ $x$ is the supremum of a
  $\emptyset^{(d)}$--computable rational sequence.
\item[d)] $\RDelta{d}\;=\;\RSigma{d}\,\cap\,\RPi{d}$.
\item[e)] $\RSigma{d}\,\cup\,\RPi{d}\;\subsetneq\;\RDelta{d+1}$.
\end{enumerate}
\end{fact}
\begin{proof}
a) \textsc{Theorem~7.8},
b+c) \textsc{Lemma~7.2},
d) \textsc{Definition~7.1},
and e) \textsc{Theorem~7.8}
in \cite{Zheng}, respectively. \qed
\end{proof}
\subsection{Type-2 Theory of Effectivity}
Specifying an encoding formalizes how to feed some general form of input
like graphs or integers into a Turing Machine with fixed alphabet $\Sigma$.
Already in the discrete case, the complexity of a problem usually depends
heavily on the chosen encoding; e.g., numbers in unary versus binary.
This dependence becomes even more important when dealing with objects
from a continuum like the set of reals and their computability.
While Recursive Analysis usually considers one particular encoding
for $\IR$, the Type-2 Theory of Effectivity (TTE) due to
\person{Weihrauch} provides
(a convenient formal framework for studying and comparing)
a variety of encodings for different universes.
Formally speaking, a \emph{representation} $\alpha$ for $\IR$
is a partial\footnote{indicated by the symbol ``$\subseteq$'',
whose absence here generally refers to total functions} 
surjective mapping $\alpha:\subseteq\Sigma^\omega\to\IR$;
and an infinite string $\bar\sigma\in\dom(\alpha)$ is
regarded as an $\alpha$--name for the real number $x=\alpha(\bar\sigma)$.

In this way, $(\alpha\to\beta)$--computing\footnote{We use this
notation instead of \cite{Weihrauch}'s $(\alpha,\beta)$--computability
to stress its connection (but not to be confused) with
$[\alpha\myto\beta]$--computability appearing in Section~\ref{secWeier}.}
a real function $f:\IR\to\IR$ means to compute a transformation
on infinite strings $F:\subseteq\Sigma^\omega\to\Sigma^\omega$
such that any $\alpha$--name $\bar\sigma$ for $x=\alpha(\bar\sigma)$
gets transformed to a $\beta$--name $\bar\tau=F(\bar\sigma)$
for $f(x)=y$, that is, satisfying $\beta(\bar\tau)=y$;
cf. \mycite{Section~3}{Weihrauch}. 
Converting
$\alpha$--names to $\beta$--names thus amounts to
$(\alpha\to\beta)$--computability of $\id:\IR\to\IR$, $x\mapsto x$,
and is called \aname{reducibility} ``$\alpha\reduceq\beta$''
\mycite{Definition~2.3.2}{Weihrauch}.
Computational \aname{equivalence}, that is mutual 
reducibility ~$\alpha\reduceq\beta$
and $\beta\reduceq\alpha$,
is denoted by ``$\alpha\equiv\beta$''
whereas ``$\alpha\reducneq\beta$'' means
$\alpha\reduceq\beta$ but $\beta\not\reduceq\alpha$.

We borrow from TTE also two ways of constructing new representations
from giving ones: The \aname{conjunction} $\alpha\myjoin\tilde\alpha$
of $\alpha$ and $\tilde\alpha$ is the least upper bound
with respect to ``\,$\reduceq\!$'' \mycite{Lemma~3.3.8}{Weihrauch};
and for (finitely or countably many) representations
$\alpha_i:\subseteq\Sigma^\omega\to A_i$, 
their \aname{product}
$\prod_i \alpha_i$ denotes a natural representation
for the set $\prod_i A_i$ \mycite{Definition~3.3.3.2}{Weihrauch}.
In particular, in order to encode $x\in\IR$ as a rational sequence
$(q_n)\in\IQ^\omega$, we (often implicitly) refer to the representation
$[\nuQ]^\omega:\subseteq\Sigma^\omega\to\IQ^\omega$ due to 
\cite[\textsc{Definition~3.1.2.4} and \textsc{Lemma~3.3.16}]{Weihrauch}.

\subsection{Arithmetic Hierarchy of Real Representations}
Observe that (the characterizations due to Fact~\ref{facRAH} of)
each level of the Real Arithmetic Hierarchy gives rise not only 
to a notion of computability for real numbers 
but also canonically to a representation for $\IR$;
for instance let

\smallskip
\noindent\begin{tabular}{l@{\,:\;\:}l}
$\pmb{\myrho}$ & encode (arbitrary!)
$x\in\IR$ as a fast convergent rational sequence $(q_n)$; \\[0.2ex]
$\pmb{\myrhol}$ & encode
$x\in\IR$ as a rational sequence $(q_n)$ with supremum $x=\sup_n q_n$; \\[0.2ex]
$\pmb{\myrho'}$ & encode 
$x\in\IR$ as a rational sequence $(q_n)$ with limit $x=\lim_n q_n$; \\[0.2ex]
$\pmb{\myrhol'}$ & encode
$x\in\IR$ as $(q_n)\subseteq\IQ$ with
$x=\mysup_i \myinf_j q_{\scriptscriptstyle\langle i,j\rangle}$; \\[0.2ex]
$\pmb{\myrho''}$ & encode
$x\in\IR$ as $(q_n)\subseteq\IQ$ with
$x=\lim_i \lim_j q_{\scriptscriptstyle\langle i,j\rangle}$.
\end{tabular}

\medskip
\noindent
As already pointed out, the first three of them are already
known and used in TTE as $\myrho$, $\myrhol$, and $\myrhon$, 
respectively \mycite{Section~4.1}{Weihrauch}.
In general one obtains, similar to Definition~\ref{defRAH},
a hierarchy of real representations as follows:
\begin{definition} \label{defRepr}
Let ~$\myrho^{(0)}:=\myrho$, ~$\myrhol^{(0)}:=\myrhol$,
~$\myrhog^{(0)}:=\myrhog$. ~ Now fix $1\leq d\in\IN$: \\
A $\myrho^{(d)}$--name for $x\in\IR$ is
(a $[\nuQ]^\omega$--name for) a rational sequence
$(q_n)$ such that%
$$\displaystyle x\;=\;
    \lim_{n_1} \; \lim_{n_2} \;\ldots \; \lim_{n_d}
      \;q_{\scriptscriptstyle\langle n_1,\ldots,n_d\rangle} 
      \enspace . $$
A $\myrhol^{(d)}$--name for $x\in\IR$ is
a (name for a) sequence
$(q_n)\subseteq\IQ$ such that
$$\displaystyle x\;=\;
    \mysup_{n_1} \; \myinf_{n_2} \;\ldots \; \mytheta_{n_{d+1}}
      \;q_{\scriptscriptstyle\langle n_1,\ldots,n_{d+1}\rangle} 
       \enspace .$$
A $\myrhog^{(d)}$--name for $x\in\IR$ is a sequence
$(q_n)\subseteq\IQ$ such that 
$\displaystyle x\;=\;
    \myinf_{n_1} \; \mysup_{n_2} \;\ldots$
\end{definition}

\noindent
Regarding Fact~\ref{facRAH}, one may see $\myrho'$ and $\myrho''$
as the first and second \emph{Jump} of $\myrho$,
respectively; same for $\myrhol'$ and $\myrhol$.

Results from \cite{Zheng} about the Real Arithmetic Hierarchy 
are easily re-phrased in terms of these representations.
Fact~\ref{facRAH}d) for example translates as follows:
\begin{quote}
$x$ is $\myrho^{(d)}$--computable ~ iff ~ it is both
$\myrhol^{(d)}$--computable and $\myrhog^{(d)}$--computable.
\end{quote}
Observe that this is a non-uniform claim whereas
closer inspection of the proofs in particular of
\textsc{Lemma~3.2} and \textsc{Lemma~3.3} in \cite{Zheng}
reveals them to hold fully uniformly so that we have
\begin{lemma} \label{lemRepr}
\;\; 
$\displaystyle\myrho\;\equiv\;\;
 \myrhol\!\myjoin\myrhog\;\;\reducneq\;\;
 \myrhol\;\;\reducneq\;\;\myrho'\;\,\equiv\;\;\,
 \myrhol'\!\myjoin\myrhog'\;\;\reducneq\;\;
 \myrhol'\;\;\reducneq\;\;\myrho''
 \;\equiv\;\ldots\;$.
\end{lemma}

\noindent
Moreover, the uniformity of \mycite{Lemma3.2}{Zheng} yields
the following interesting
\begin{scholiumf}\footnotetext{A \aname{scholium} is ``\emph{a 
note amplifying a proof or course of reasoning,
as in mathematics}'' \cite{Dictionary}} \label{schLiminf}
Let $\tildemyrhol'$ denote the representation encoding
$x\in\IR$ as $(q_n)\subseteq\IQ$ with $x=\liminf_n q_n$;
and $\hatmyrhol'$ similarly with the additional
requirement that $q_n<x$ for infinitely many $n$. \\
Then it holds ~$\hatmyrhol'\equiv\tildemyrhol'\equiv\myrhol'$~
($\hatmyrhol'\reduceq\tildemyrhol'\reduceq\,\myrhol'$~
being the trivial direction).
\end{scholiumf}
\section{Computability and Continuity} \label{secContinuous}
Recursive Analysis has established as folklore that any
computable real function is continuous.
More precisely, computability of a partial function
from/to infinite strings $f:\subseteq\Sigma^\omega\to\Sigma^\omega$
requires continuity with respect to the Cantor Topology $\tau_{\text{C}}$
\mycite{Theorem~2.2.3}{Weihrauch};
and this requirement carries over to functions
$f:\subseteq A\to B$ on other topological spaces
$(A,\tau_A)$ and $(B,\tau_B)$
where input $a\in A$ and output $b=f(a)$
are encoded by respective \emph{admissible} representations
$\alpha$ and $\beta$. Roughly speaking, this
property expresses that the mappings
$\alpha:\subseteq\Sigma^\omega\to A$ and
$\beta:\subseteq\Sigma^\omega\to B$
satisfy a certain compatibility condition
with respect to the topologies $\tau_A$/$\tau_B$
and $\tau_{\text{C}}$ involved.
For $A=B=\IR$, the (standard) representation
$\myrho$ for example is admissible \mycite{Lemma~4.1.4.1}{Weihrauch},
thus recovering the folklore claim.

Now in order to treat and non-trivially investigate computability
also of discontinuous real functions $f:\IR\to\IR$,
there are basically two ways out:
Either enhance the underlying Type-2 Machine model
or resort to non-admissible representations.
It turns out that for either choice, at least the
straight-forward approaches fail: \\ \noindent
\textbullet ~ extending Turing Machines with oracles
\quad as well as \\ \noindent
\textbullet ~ considering weakened representations for $\IR$.
\subsection{Type-2 Oracle Computation} \label{secOracles}
Specifically concerning the first approach,
most results in Computable Analysis relativize.
Specifically we make 
\begin{observation} \label{obsRelative}
Let $\calO\subseteq\Sigma^*$ be arbitrary.
Replace in {\rm\mycite{Definition~2.1.1}{Weihrauch}}
the Turing Machine $\calM$ by $\calM^\calO$, that is, 
one with oracle access to $\calO$. This
Type-2 Computability \emph{in $\calO$} still satisfies%
\begin{enumerate}
\item[a)]
  closure under composition
  {\rm\mycite{Theorem~2.1.12}{Weihrauch}};
\item[b)]
  computability of string functions requires continuity
  {\rm\mycite{Theorem~2.2.3}{Weihrauch}};
\item[c)]
  same for computable functions on represented spaces
  with respect to \emph{admissible} representations 
  {\rm\mycite{Corollary~3.2.12}{Weihrauch}}.
\end{enumerate}
In particular, the Main Theorem of Recursive Analysis
{\rm\mycite{Theorem~4.3.1}{Weihrauch}}
relativizes.
\end{observation}
A strengthening and counterpart to 
Observation~\ref{obsRelative}b), we have
\begin{lemma} \label{lemRelative}
For a partial function on infinite strings
~$f:\subseteq\Sigma^\omega\to\Sigma^\omega$,
the following are equivalent:
\begin{itemize}
\item[\textbullet] There exists an oracle $\calO$
such that $f$ is computable relative to $\calO$;
\item[\textbullet] $f$ is Cantor-continuous 
and $\dom(f)$ is a $\Gdelta$--set.
\end{itemize}
\end{lemma}
Compare this with Type-1 Theory (that is,
computability on finite strings) where every function
$f:\subseteq\Sigma^*\to\Sigma^*$ is recursive
in some appropriate $\calO\subseteq\Sigma^*$.
\begin{proof}[Lemma~\ref{lemRelative}]
If $f$ is recursive in $\calO$, then it is also continuous
by Observation~\ref{obsRelative}b), that is,
the relativized version of \mycite{Theorem~2.2.3}{Weihrauch}.
Furthermore the relativization of \mycite{Theorem~2.2.4}{Weihrauch}
reveals $\dom(f)$ to be a $\Gdelta$--set.

Conversely suppose that continuous $f$ has $\Gdelta$ domain.
Then $f=h_\omega$ for some monotone total function
$h:\Sigma^*\to\Sigma^*$ according to
\mycite{Theorem~2.3.7.2}{Weihrauch} where, by
\mycite{Definition~2.1.10.2}{Weihrauch},
$h_\omega:\subseteq\Sigma^\omega\ni\bar\sigma\mapsto
\sup_{n} h(\sigma_1\ldots\sigma_n)$ denotes the (existing
and unique) extension of $h$ from $\Sigma^*$ to $\subseteq\Sigma^\omega$.
A classical Type-1 function on finite strings, this $h$
is recursive in a certain oracle $\calO\subseteq\Sigma^*$.
The relativization of \mycite{Lemma~2.1.11.2}{Weihrauch}
then asserts also $h_\omega=f$ to be computable in $\calO$.
\qed
\end{proof}
The conclusion of this subsection is that 
oracles do not increase the computational power
of a Type-2 Machine sufficiently in order to handle
also discontinuous functions. So let us proceed
to the second approach to real hypercomputation:
\subsection{Weaker Representations for Reals} \label{secWeak}
In the present section we are interested in
relaxations of the standard representation $\myrho$
for single reals
and their effect on the computability of function
evaluation $x\mapsto f(x)$.
Since, with exception of $\myrho$, none of the
ones introduced in Definition~\ref{defRepr} is admissible
with respect to the usual Euclidean\footnote{%
it might be admissible w.r.t. some other, typically artificial
topology, though} topology on $\IR$
\mycite{Lemma~4.1.4, Example~4.1.14.1}{Weihrauch},
the relativized Main Theorem (Observation~\ref{obsRelative}c)
is not applicable. Hence,
chances are good for evaluation $x\mapsto f(x)$ to become computable
even for discontinuous $f:\IR\to\IR$; and indeed we have the following
\\[1ex]\begin{minipage}[c]{60ex}
\begin{example} \label{exHeaviside}
\person{Heaviside}'s function \\[0.7ex]
\centerline{$\displaystyle\Heavi:\IR\to\IR, \qquad x\mapsto 0 \text{ ~for~ }x\leq0,
\quad x\mapsto 1 \text{ ~for~ }x>0$}
\\[0.5ex]
is both $(\myrhol\to\myrhol)$--computable
and $(\myrhol'\to\myrhol')$--computable.%
\end{example}
\end{minipage}\hfill\begin{minipage}[c]{22ex}
\includegraphics[width=\textwidth]{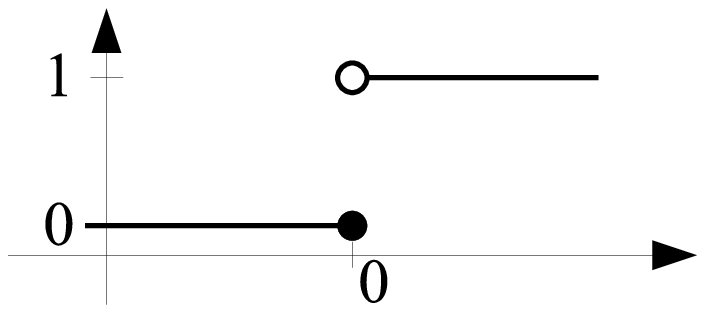}%
\end{minipage}
\begin{proof}
Given $(q_n)\subseteq\IQ$ with $x=\sup_n q_n$,
exploit $(\nuQ\to\nuQ)$--computability
of the restriction $\Heavi|_{\IQ}:\IQ\to\{0,1\}$ to
obtain  $p_n:=\Heavi(q_n)$. Then indeed, $(p_n)\subseteq\IQ$ has
$\sup_n p_n=\Heavi(x)$: In case $x\leq0$, $q_n\leq0$ and hence
$p_n=0$ for all $n$; whereas in case $x>0$, $q_n>0$
and hence $p_n=1$ for some $n$.

Let $x\in\IR$ be given by a rational double sequence
$(q_{i,j})$ with $x=\mysup_i\myinf_j q_{i,j}$.
Proceeding from $q_{i,j}$ to
$\tilde q_{i,j}:=\max\{q_{0,j},\ldots,q_{i,j}\}$,
we assert $\myinf_j\tilde q_{i+1,j}\geq\myinf_j\tilde q_{i,j}$.
Now compute $p_{i,j}:=\Heavi(\tilde q_{i,j}-2^{-i})$. Then in case $x\leq0$,
it holds
$\forall i \exists j: \tilde q_{i,j}\leq 2^{-i}$,
i.e., $p_{i,j}=0$
and thus $\mysup_i\myinf_j p_{i,j}=0=\Heavi(x)$. Similarly in case
$x>0$, there is some $i_0$ such that
$\myinf_j \tilde q_{i_0,j}>x/2$ and thus
$\myinf_j \tilde q_{i,j}>x/2$ for all $i\geq i_0$.
For $i\geq i_0$ with $2^{-i}\leq x/2$,
it follows $p_{i,j}=1$ $\forall j$
and therefore $\mysup_i\myinf_j p_{i,j}=1=\Heavi(x)$.
\qed
\end{proof}
So real function hypercomputation based on weaker representations
indeed does allow for effective evaluation of some discontinuous
functions. On the other hand, they still impose well-known
topological restrictions:
\begin{fact} \label{facFolklore}
Consider $f:\IR\to\IR$.%
\begin{itemize}
\item[a)]
 If $f$ is ~$(\myrho\to\myrho)$--computable,
 then it is continuous.
\item[b)]
 If $f$ is ~$(\myrho\to\myrhol)$--computable,
 then it is lower semi-continuous.
\item[c)]
 If $f$ is ~$(\myrhol\to\myrhol)$--computable,
 then it is monotonically 
 increasing.
\item[d)]
 If $f$ is ~$(\myrho'\to\myrho')$--computable,
 then it is continuous.
\end{itemize}
The claims remain valid under oracle-supported computation.
\end{fact}
Claim~a) is the Main Theorem.
For b) see \cite{SemiTCS} and recall,
e.g. from \mycite{Chapter~6.7}{Analysis},
that $f:\IR\to\IR$ is lower semi-continuous iff
$f(\lim_n x_n)\leq\liminf_n f(x_n)$
for all convergent sequences $(x_n)$;
equivalently: $f^{-1}\big[(y,\infty)\big]$
is open for any $y\in\IR$.
The establishing of d) in \mycite{Section~6}{Naive}
caused some surprise.
We briefly sketch the according proofs as a preparation
for those of Theorem~\ref{thNecessary} below.
\begin{proof}
\begin{enumerate}
\item[a)] Suppose for a start that Heaviside's function,
in spite of its discontinuity at $x=0$, be $(\myrho\to\myrho)$--computable
by some Type-2 Machine $\calM$.
Feed the rational sequence $q_n:=2^{-n}$,
a valid $\myrho$--name for $x$, to this $\calM$.
By presumption it will then spit out a sequence
$(p_m)_{_m}\subseteq\IQ$ with $|p_m-y|\leq 2^{-m}$ for $y=\Heavi(x)=0$;
in particular, $|p_2-\tilde y|>2^{-2}$ for $\tilde y:=1$.
Up to output of $p_2$, $\calM$ has executed a finite number $N\in\IN$ of
operations and in particular read at most the initial part
$p_0,p_1,\ldots,p_N$ of the input.

Now re-use $\calM$ in order to evaluate $\Heavi$ at $\tilde x:=p_N>0$
$\myrho$--encoded as the rational sequence
$(\tilde q_n):=(q_0,q_1,\ldots,q_N,q_N,\ldots)$
coinciding with $(p_n)$ for $n\leq N$.
Being a deterministic machine, $\calM$ will then proceed
exactly as before for its first $N$ steps;
in particular the output $(\tilde p_m)$ agrees with $(p_m)$ up to $m=2$.
Hence $|\tilde p_2-\tilde y|>2^{-2}$
contradicting that $\calM$ is supposed to output a
$\myrho$--name for $\tilde y=\Heavi(\tilde x)$.

\smallskip
For the case of a general function $f:\IR\to\IR$ with
discontinuity at some $x\in\IR$, let
$y=f(x)\not=\lim_k f(x_k)=\tilde y$ with a
real sequence $x_k$ converging to $x$.
There exists $M\in\IN$ with $|y-\tilde y|>2^{-M+2}$;
by possibly proceeding to an appropriate subsequence of $(x_k)$,
we may suppose w.l.o.g. that $|x-x_k|\leq2^{-k-2}$
and $|f(x_k)-\tilde y|\leq2^{-M}$.
Then there is a rational double sequence $(q_{k,n})$ such that
$|x_k-q_{k,n}|\leq2^{-n-1}$; thus $|x-q_{n,n}|\leq2^{-n}$.
We may therefore feed $(q_{n,n})$ as a $\myrho$--name in order
to evaluate $f$ at $x$ and obtain in turn a $\myrho$--name
$(p_m)\subseteq\IQ$ for $y$.
As before, $p_M$ is output after having only read
some finite initial part $(q_{n,n})_{_{n\leq N}}$ of the
input. Then
$$|q_{n,n}-x_N|\leq|q_{n,n}-x_n|+|x_n-x|+|x-x_N|
\leq 2^{-n-1}+2^{-n-2}+2^{-N-2}\leq2^{-n}$$
for $n\leq N$
reveals this very initial part to also be the start
of a valid $\myrho$--name for $\tilde x:=x_N$ whereas
$$2^{-M+2}<|y-\tilde y|
\leq |y-p_M|+|p_M-f(\tilde x)|+|f(\tilde x)-\tilde y|
\leq 2^{-M} + |p_M-f(\tilde x)|+ 2^{-M}$$
shows 
that $(p_m)_{m\leq M}$ is not a valid
initial part of a $\myrho$--name for $f(\tilde x)$: contradiction.
\item[b)]
We prove $(\myrho\to\myrhol)$--uncomputability of the flipped
Heaviside Function%
\[\barH:0\geq x\mapsto 1, \qquad 0<x\mapsto 0 \qquad\qquad
\smash{\raisebox{-5ex}{\includegraphics[height=9ex]{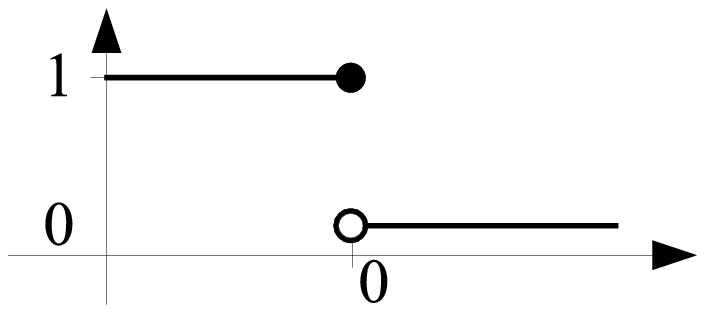}}} \]
as a prototype lacking lower semi-continuity.
\\
Consider again the $\myrho$--name $q_n:=2^{-n}$ for
$x=0$ which the hypothetical Type-2 Machine transforms
into a $\myrhol$--name for $y=\barH(x)=1$, that is,
a sequence $(p_m)\subseteq\IQ$ with $\sup_m p_m=y$;
In particular $p_M\geq\tfrac{2}{3}$ for some $M\in\IN$
gets output
having read only $(q_n)_{_{n\leq N}}$ for some $N\in\IN$.
The latter finite segment is also the initial part of
a valid $\myrho$--name for $\tilde x=q_N>0$ whereas
$(p_m)_{_{m\leq M}}$ has $\sup\geq\tfrac{2}{3}$ and
thus is not the initial part of a valid $\myrhol$--name
for $\tilde y=\barH(\tilde x)=0$: contradiction.

This proof for the case $\barH$ carries over to
an arbitrary $f:\IR\to\IR$
just like in a), that is, by replacing $q_n=2^{-n}$
with rational approximations to a general sequence
$x_n\in\IR$ witnessing violated lower semi-continuity
of $f$ in that $f(\lim_n x_n)>\liminf_n f(x_n)$.
\item[c)]
As in a) and b), we treat for notational simplicity
the case of $f:\IR\to\IR$ violating
monotonicity in that $f(0)=1$ and $f(1)=0$;
the general case can again be handled similarly.
Feed the $\myrhol$--name $(q_n)=(0,0,\ldots)$ for $x=0$
into a machine which be presumption produces a sequence
$(p_m)\subseteq\IQ$ with $\sup p_m=1$ and in particular
$p_M\geq\tfrac{2}{3}$ for some $M\in\IN$.
Up to output of $p_M$, only $(q_n)_{_{n\leq N}}$ has
been read for some $N\in\IN$. Now consider
the rational sequence $(\tilde q_n)$ consisting
of $N$ zeros followed by an infinity of 1s,
that is, a valid $\myrhol$--name for $\tilde x=1$.
This new input will cause the machine to output
a sequence $(\tilde p_m)\subseteq\IQ$ coinciding
with $(p_m)$ for $m\leq M$; in particular
$\tilde p_M\geq\tfrac{2}{3}$ contradicting that
$(\tilde p_m)$ is supposed to satisfy
$\sup_m\tilde p_m=f(\tilde x)=0$.
\item[d)]
Suppose that, in spite of its discontinuity at $x=0$,
$\barH$ be $(\myrho'\to\myrho')$--computable by some
Type-2 Machine $\calM$.

Consider the sequence $q^{(1)}:=\big(q^{(1)}_n\big)\subseteq\IQ$,
$q^{(1)}_n:\equiv1$, which is by definition a valid $\myrho'$--name
for $1=:x^{(1)}=\lim_n q^{(1)}_n$.
So upon input of $q^{(1)}$, $\calM$ will generate
a corresponding sequence $p^{(1)}\subseteq\IQ$ as a
$\myrho'$--name for $y^{(1)}=\barH\big(x^{(1)}\big)=0$,
that is, satisfying $\lim_m p^{(1)}_m=0$; in particular,
$p^{(1)}_{m_1}\leq\tfrac{1}{3}$ for some $m_1\in\IN$.
Up to this output, $\calM$ has read only a finite initial part
of the input $q^{(1)}$, say, up to $n\leq n_1$.

Next consider the sequence $q^{(2)}\subseteq\IQ$ defined by
$q^{(2)}_n:=1$ for $n\leq n_1$ and $q^{(2)}_n:=\tfrac{1}{2}$
for $n>n_1$: a valid $\myrho'$--name for $x^{(2)}=\tfrac{1}{2}$
which $\calM$ by presumption transforms into a sequence $p^{(2)}\subseteq\IQ$
with $\lim_m p^{(2)}_m=y^{(2)}=\barH\big(x^{(2)}\big)=0$;
in particular, $q^{(2)}_{m_2}\leq\tfrac{1}{3}$ for some
$m_2>m_1$. However, due to $\calM$'s deterministic behavior
and since $q^{(1)}$ and $q^{(2)}$ initially coincide,
it still holds $p^{(2)}_{m_1}\leq\tfrac{1}{3}$.

Now by repeating the above argument we obtain a sequence
of sequences $q^{(k)}\subseteq\IQ$, each constant for $n\geq n_k$
of value (and thus a valid $\myrho'$--name for) $x^{(k)}=2^{-k+1}$
and transformed by $\calM$ into a sequence $p^{(k)}\subseteq\IQ$
satisfying $p^{(k)}_{m_i}\leq\tfrac{1}{3}$ for $i=1,\ldots k$
with strictly increasing $(n_k),(m_k)\subseteq\IN$.
The ultimate sequence $q^{(\omega)}\subseteq\IQ$, well-defined by
$q^{(\omega)}_n:=q^{(k)}_n$ for $n\leq n_k$
(and in fact the limit of the sequence of sequences
$\big(q^{(k)}\big)_{_k}$ with respect to Baire's Topology),
therefore converges
to (and is hence a valid $\myrho'$--name for) $x^{(\omega)}=0$;
and it gets mapped by $\calM$ to a sequence $q^{(\omega)}\subseteq\IQ$
satisfying $q^{(\omega)}_m\leq\tfrac{1}{3}$ for infinitely
many $m$ contradicting that a valid $\myrho'$--name
for $y^{(\omega)}=\barH\big(x^{(\omega)}\big)=1$
should have $\lim_m=1$.
\end{enumerate}
\noindent
Being only information-theoretic,
the above arguments obviously relativize.%
\qed\end{proof}
The main result of the present section is an extension
of Fact~\ref{facFolklore}
to one level up on the hierarchy of real representations
from Definition~\ref{defRepr}. This suggests similar
claims to hold for the entire hierarchy and might not
be as surprising any more as Fact~\ref{facFolklore}d) in \cite{Naive};
nevertheless, already this additional step makes
proofs significantly more involved.
\begin{theorem}[First Main Theorem of Real Hypercomputation]
\label{thNecessary} ~ \\ Consider $f:\IR\to\IR$.
\begin{itemize}
\item[a)]
  If $f$ is ~$(\myrho'\to\myrhol')$--computable,
  then it is lower semi-continuous.
\item[b)]
  If $f$ is ~$(\myrhol'\to\myrhol')$--computable,
  then it is monotonically increasing.
\item[c)]
  If $f$ is ~$(\myrho''\to\myrho'')$--computable,
  then it is continuous.
\end{itemize}
The claims remain valid under oracle-supported computation.
\end{theorem}
We point out that the proofs of Fact~\ref{facFolklore} proceed
by constructing an input for which a presumed machine $\calM$ fails to
produce the correct output. They differ however in the `length'
of these constructions: for Claims~a) to c), the counter-example
inputs are obtained by running $\calM$ for a finite number of steps
on a single, fixed argument; whereas in the proof of Claim~d),
$\calM$ is repeatedly started on an adaptively extended sequence of
arguments. The latter argument may thus be considered as of
length $\omega$, the first infinite ordinal.
Our proof of Theorem~\ref{thNecessary}c) will be even longer
and is therefore put into the following subsection.
\subsection{Proof of Theorem~\ref{thNecessary}} \label{secProof}
As in the proof of Fact~\ref{facFolklore},
we treat the special case of the flipped
Heaviside Function $\barH$ for reasons of
notational convenience and clarity of
presentation; the according arguments
can be immediately extended to the general case.
\begin{myclaim} \label{clOne}
$\barH:\IR\to\IR$ is not $(\myrho'\to\myrhol')$--computable.
\end{myclaim}
\begin{proof}
Suppose a Type-2 Machine $\calM$
$(\myrho'\to\myrhol')$--computes $\barH$.
In particular, upon input of $x^{(1)}=1$ in form of
the sequence $q^{(1)}=(q^{(1)}_n)$ with $q^{(1)}_n:\equiv1$, $\calM$ will output
a rational double sequence $p^{(1)}=(p^{(1)}_{k,\ell})$ with
$0=y^{(1)}:=\barH(x^{(1)})=
\mysup\limits_k\myinf\limits_\ell p^{(1)}_{k,\ell}$.
Observe that $p^{(1)}_{1,\ell_1}\leq\tfrac{1}{3}$ for some
$\ell_1$. When writing $p^{(1)}_{1,\ell_1}$, $\calM$ has only
read a finite part of $(q^{(1)}_n)$, say, up to $n_1$.

Now consider $x^{(2)}:=\tfrac{1}{2}$, given by way of
the sequence $q^{(2)}$ with
\quad $q^{(2)}_n:=1$ ~ for ~ $n<n_1$ \quad and\quad
$q^{(2)}_n:=\tfrac{1}{2}$ ~ for ~ $n\geq n_1$.
Then, too, $\calM$ will output a double sequence $p^{(2)}$
with $0=y^{(2)}=\mysup\limits_k\myinf\limits_\ell p^{(2)}_{k,\ell}$.
Observe that, similarly, some $p^{(2)}_{2,\ell_2}\leq\tfrac{1}{3}$
is output having read only a finite part of $(q^{(2)}_n)$,
say, up to $n_2$. Moreover, as
$q^{(1)}$ and $q^{(2)}$ coincide up to $n_1$
and since $\calM$ operates deterministically,
$p^{(2)}_{1,\ell_1}=p^{(1)}_{1,\ell_1}\leq\tfrac{1}{3}$.
\begin{figure}[htb]
\includegraphics[width=0.99\textwidth]{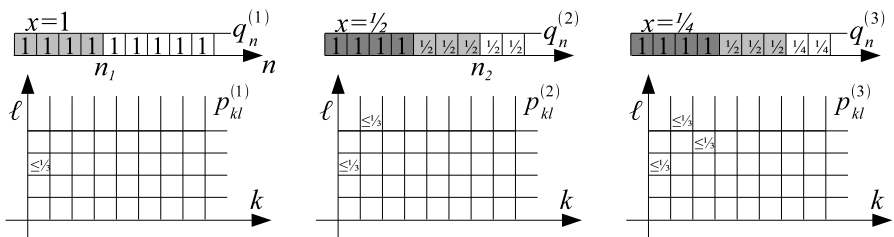}\vspace*{-4ex}%
\caption{\label{figLowersemi}Illustration to the
iterative construction employed in the proof of
Claim~\ref{clOne}}%
\end{figure}

Continuing this process with $x^{(k)}:=2^{-k+1}$
for $k=3,4,\ldots$ as
indicated in Figure~\ref{figLowersemi} eventually yields a
rational sequence $q^{(\omega)}$ with
$\lim_n q^{(\omega)}_n=:x^{(\omega)}=0$, upon input of which $\calM$ outputs
a double sequence $p^{(\omega)}$
such that $p^{(\omega)}_{k,\ell_k}\leq\tfrac{1}{3}$
for all $k=1,2,\ldots$. In particular,
$y^{(\omega)}:=
\mysup\limits_k\myinf\limits_\ell p^{(\omega)}_{k,\ell}\leq\tfrac{1}{3}$
whereas $\barH(x^{(\omega)})=1$ : contradiction.
\qed\end{proof}
Notice that the above proof involves one-dimensionally
indexed sequences $(q_n)$ for input and two-dimensionally
indexed ones $(p_{k,\ell})$ for output.
We now proceed a step further in proof difficulty,
namely involving two-dimensional indices for both
input and output in order to establish Item~b).
\begin{myclaim} \label{clTwo}
Let $f:\IR\to\IR$ violate monotonicity
in that $f(0)=1$ and $f(1)=0$.
Then, $f$ is not $(\myrhol'\to\myrhol')$--computable.
\end{myclaim}
\begin{proof}
 We construct a $\myrhol'$--name for $x=0$
 from an iteratively defined sequence of initial segments of
 $\myrhol'$--names for $x=1$:

 Start with $q^{(1)}_{i,j}:=1$ for all $i,j$. Then,
 $q^{(1)}=(q^{(1)}_{i,j})$ is obviously a $\myrhol'$--name for $x=1$
 and thus yields by presumption, upon input to $\calM$, a
 $\myrhol'$--name $p^{(1)}_{k,\ell}$ for $f(1)=0$,
 that is, with $0=\mysup\limits_k\myinf\limits_{\ell} p^{(1)}_{k,\ell}$.
 In particular, $p^{(1)}_{1,\ell_1}\leq\tfrac{1}{3}$ for
 some $\ell_1$.
\begin{figure}[htb]
\includegraphics[width=0.99\textwidth]{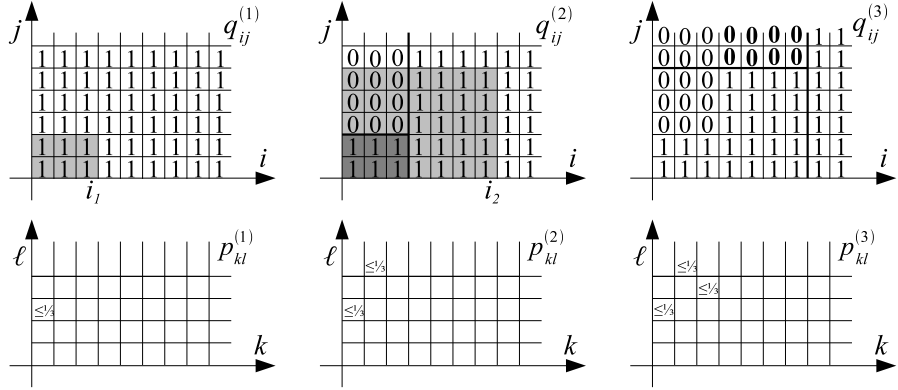}\vspace*{-4ex}%
\caption{\label{figMonoton}Illustration to the
iterative construction employed in the proof of
Claim~\ref{clTwo}}%
\end{figure}

Until output of $p^{(1)}_{1,\ell_1}$, $\calM$
 has read only finitely many entries of $q^{(1)}$;
 say, up to $i_1$ and $j_1$, that is, covered
 in Figure~\ref{figMonoton} by the light gray rectangle.
 Now consider $q^{(2)}$ defined as in this figure:
 Since $\myinf_j q^{(2)}_{i,j}=0$ for $i\leq i_1$ and
 $\myinf_j q^{(2)}_{i,j}=1$ for $i>i_1$,
 $\mysup\limits_i\myinf\limits_j q^{(2)}_{i,j}=1$, that is, this is again
 valid $\myrhol'$--name for $x=1$; and again, $\calM$ will
 by presumption convert $q^{(2)}$ into a $\myrhol'$--name
 $p^{(2)}$ for $f(1)=0$. In particular,
 $p^{(2)}_{2,\ell_2}\leq\tfrac{1}{3}$ for
 some $\ell_2$; and, being a deterministic machine, $\calM$'s
 operation on the initial part (dark gray) on which input $q^{(2)}$
 coincides with input $q^{(1)}$ will first have generated
 the same initial output, namely
 $p^{(2)}_{1,\ell_1}=p^{(1)}_{1,\ell_1}\leq\tfrac{1}{2}$.
\\
 Again, until output of $p^{(2)}_{2,\ell_2}$, $\calM$
 has read only a finite part of $q^{(2)}$ of, say, up to $i_2>i_1$
 (light gray). By now considering input $q^{(3)}$ with
 $\myinf_j q^{(3)}_{i,j}=0$ for $i\leq i_2$ as in
 Figure~\ref{figMonoton}, we arrive at $p^{(3)}$ and $\ell_3$ with
 $p^{(3)}_{1,\ell_1},p^{(3)}_{2,\ell_2},p^{(3)}_{3,\ell_3}\leq\tfrac{1}{3}$;
 and so on with $i_3, q^{(4)}, p^{(4)}, \ell_4, i_4$, \ldots
\\
 Finally observe that continuing these arguments eventually
 leads  to a rational double sequence
 $q^{(\omega)}=(q^{(\omega)}_{i,j})$ which has
 $\myinf_j q^{(\infty)}_{i,j}=0$ for $i\leq i_{\infty}=\infty$%
---and is therefore a valid $\myrhol'$--name for $x=0$ (rather
 than $x=1$)---but gets mapped by $\calM$ to
 $p^{(\omega)}=(p^{(\omega)}_{k,\ell})$ with
 $\myinf_\ell p^{(\infty)}_{k,\ell}
   \leq p^{(\infty)}_{k,\ell_k}\leq\tfrac{1}{3}$ for all $k$.
 Since $f(0)=1$, this contradicts our presumption that
 $\calM$ maps $\myrhol'$--names for $x$
 to $\myrhol'$--names for $f(x)$.
\qed\end{proof}
The above proofs involving
$\myrho'$ and $\myrhol'$
proceeded by constructing an infinite sequence of inputs
$q^{(1)},q^{(2)},\ldots,q^{(\omega)}$~
(each possibly a multi-indexed sequence of its own).
For finally asserting Claim~c) involving $\myrho''$,
we will extend this method from length $\omega$,
the first infinite ordinal, to an even longer one.
\begin{myclaim} \label{clThree}
$\barH:\IR\to\IR$
is not $(\myrho''\to\myrho'')$--computable.
\end{myclaim}
\begin{proof}
Outwit a Type-2 Machine $\calM$, presumed to realize
this computation, as follows:%
\begin{enumerate}
\item[i)]
Take $q^{(1)}$ to be the constant double sequence 1,
i.e., $q^{(1)}_{i,j}:=1$ for all $i,j$.
Being a $\myrho''$--name for $1$, it is by
presumption mapped to a $\myrho''$--name
$p^{(1)}$ for $\barH(1)=0$, that is, satisfying
$\lim_k\lim_\ell p^{(1)}_{k,\ell}=0$. In particular,
almost every column $\#k$ contains an entry $\#\ell$
with $p^{(1)}_{k,\ell}\leq\tfrac{1}{3}$. Until output of
the first such $p^{(1)}_{k_1,\ell_1}$, $\calM$ has
read only a finite part of $q^{(1)}$---say,
up to $i_1,j_1$.

\begin{figure}[htb]
\includegraphics[width=0.99\textwidth]{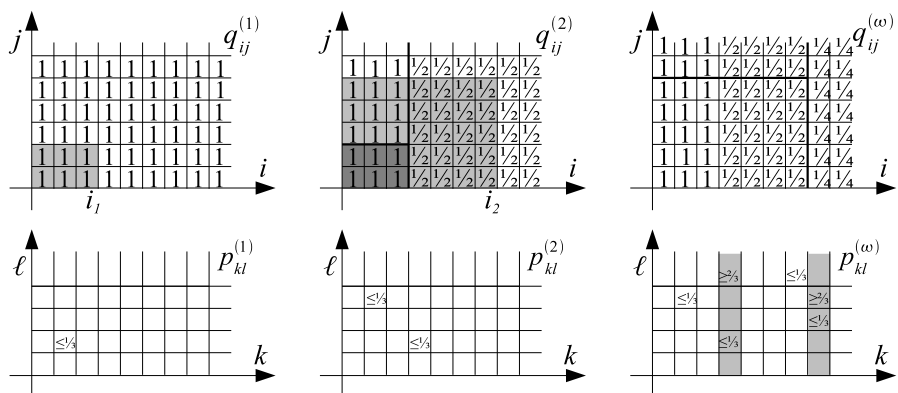}\vspace*{-4ex}%
\caption{\label{figOmega1}The first infinitely long
iterative construction employed in the proof of
Claim~\ref{clThree}}%
\end{figure}

\item[ii)]
Observe that this Argument~i) equally applies to the
scaled input sequence $2^{-m}\cdot q^{(1)}$ for any $m$.
So define $q^{(2)}_{i,j}:=q^{(1)}_{i,j}$ for
$j\leq j_1$ (i.e., inherit the initial part of $q^{(1)}$)
and $q^{(2)}_{i,j}:\equiv\tfrac{1}{2}$ for $j>j_1$.
Now upon input of this $q^{(2)}$,
$\calM$ will output $p^{(2)}$ with, again,
infinitely many $p^{(2)}_{k,\ell}\leq\tfrac{1}{3}$,
the first one---$(k_2,\ell_2)$, say---after having
read $q^{(2)}$ only up to some $(i_2,j_2)$. Furthermore
$\calM$'s determinism implies
$p^{(2)}_{k_1,\ell_1}=p^{(1)}_{k_1,\ell_1}\leq\tfrac{1}{3}$.

By repeating for $m=2,3,\ldots$, we eventually obtain%
---similarly to the proof of Claim~\ref{clTwo}---an input
sequence $q^{(\omega)}$ with $q^{(\omega)}_{i,j}$
with $\lim_i\lim_j q^{(\omega)}_{i,j}=0$, that is,
a valid $\myrho''$--name for $x=0$ (rather than 1).
This is mapped by $\calM$ to $p^{(\omega)}$ with
$p^{(\omega)}_{k_m,\ell_m}\leq\tfrac{1}{3}$ for all $m$.
On the other hand, $p^{(\omega)}$ is by presumption a
$\myrho''$--name for $\barH(0)=1$.
Therefore, there are infinitely many $m$
with $p^{(\omega)}_{m,\ell}\geq\tfrac{2}{3}$
for some $\ell>\ell_m$ \underline{and}
$p^{(\omega)}_{m,\ell_m}\leq\tfrac{1}{3}$;
see the grey columns in the right part of
Figure~\ref{figOmega1}.
\item[iii)]
Since this gives no contradiction yet, we proceed
by considering the first such column $m$ containing
an entry $\leq\tfrac{1}{3}$ as well as an entry
$\geq\tfrac{2}{3}$. Take the initial
part of the input $q^{(\omega)}$ ---
up to $(i_{\omega},j_{\omega})$, say,
depicted in grey in the left part of Figure~\ref{figOmega2} ---
that $\calM$ has read until output of both of them;
extend it with $\tfrac{1}{2}$s in top direction
and with $1$s to the right. Feed this $\myrho''$--name
for $x=1$ into $\calM$ until output of an entry
$p_{k,\ell}\leq\tfrac{1}{3}$
in some column $k$ beyond $m$. Then repeat extending
to the right with $1$s replaced by $\tfrac{1}{2}$s
for a second entry $p_{k,\ell}\leq\tfrac{1}{3}$.

\begin{figure}[htb]
\includegraphics[width=0.99\textwidth]{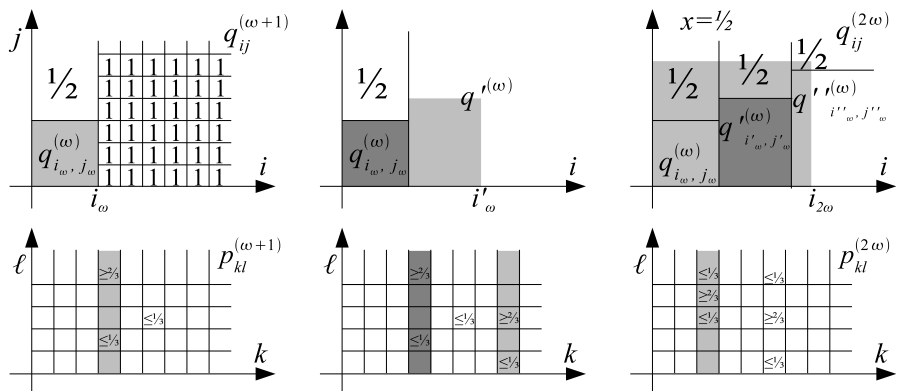}\vspace*{-4ex}%
\caption{\label{figOmega2}Second infinitely long
iterative construction employed in the proof of
Claim~\ref{clThree}}%
\end{figure}

More generally, proceed similarly as in ii)
and extend $q^{(\omega)}_{i_{\omega},i_{\omega}}$
to the right in such a way with some $\myrho''$--name
$q'^{(\omega)}$ for $x=0$
as to obtain another column $m'$
with both entries $\leq\tfrac{1}{3}$ and
$\geq\tfrac{2}{3}$; see the middle part of Figure~\ref{figOmega2}.
Again, $\calM$ outputs the latter two entries having
read only a finite part; say, up to $(i'_\omega,j'_\omega)$.

Now extend this part, too, with $\tfrac{1}{2}$ in
top direction and with another $q''^{(\omega)}$
obtained, again, as in ii) for a third column $m''$
with both entries $\leq\tfrac{1}{3}$ and
$\geq\tfrac{2}{3}$; and so on.

This eventually
leads to an input $q^{(2\omega)}$ which,
due to the extensions to the top, represents
a $\myrho''$--name for $x=\tfrac{1}{2}$ and is
thus mapped by presumption to a $\myrho''$--name
$p^{(2\omega)}$ for $\barH(\tfrac{1}{2})=0$.
In particular, almost every column of $p^{(2\omega)}$
has almost every entry $\leq\tfrac{1}{3}$ while maintaining
infinitely many columns with preceding entries
$\leq\tfrac{1}{3}$ and $\geq\tfrac{2}{3}$;
see the right part of Figure~\ref{figOmega2}. This asserts the existence
of infinitely many columns in $p^{(2\omega)}$
containing $\leq\tfrac{1}{3}$, $\geq\tfrac{2}{3}$, and
$\leq\tfrac{1}{3}$ in order.
And again, already a finite initial part
of $q^{(2\omega)}$ up to some $(i_{2\omega},j_{2\omega})$
gives rise to the first such triple.
\item[iv)]
Notice that the arguments in iii)
similarly yield the existence of an appropriate,
scaled counter-part $\tfrac{1}{2}q'^{(2\omega)}$ of
$q^{(2\omega)}$, of some $\tfrac{1}{4}q''^{(2\omega)}$,
and so on, all leading to output containing infinitely
many columns with alternating triples as above.
We now construct input $q^{(3\omega)}$ leading to
output $p^{(3\omega)}$ containing an infinity of
columns, each with four entries
$\leq\tfrac{1}{3}$, $\geq\tfrac{2}{3}$,
$\leq\tfrac{1}{3}$, \underline{and} $\geq\tfrac{2}{3}$.

To this end, take the initial part of $q^{(2\omega)}$
leading to output of the first column with alternating
triple in the above sense; then extend it with the
initial part of the scaled version $\tfrac{1}{2}q'^{(2\omega)}$
leading to another column with such a triple; and so on.
Observing that, due to the scaling, the thus obtained
$q^{(3\omega)}$ represents a $\myrho''$--name for
$x=0$, almost every column of the output $p^{(3\omega)}$
representing $\barH(0)=1$ contains entries $\geq\tfrac{2}{3}$
in addition to the infinitely many columns with triples as
above; see the left part of Figure~\ref{figOmega3}.

\begin{figure}[htb]
\includegraphics[width=0.99\textwidth]{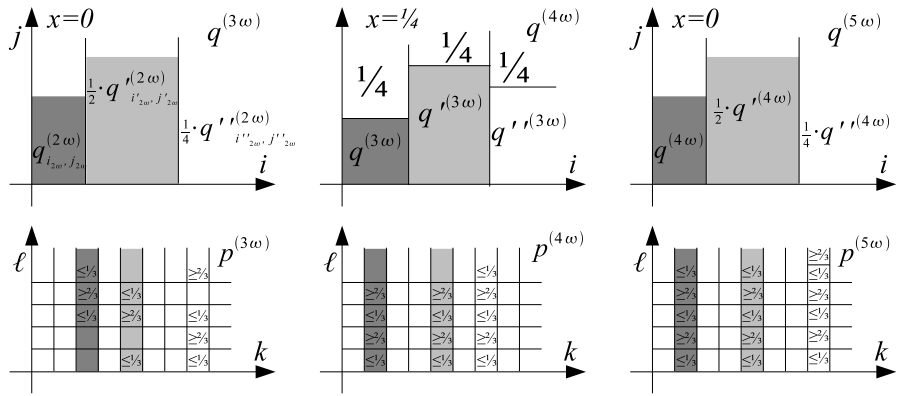}\vspace*{-4ex}%
\caption{\label{figOmega3}Third, fourth, and fifth 
infinitely long iterative construction employed in the proof of
Claim~\ref{clThree}}%
\end{figure}
\item[v)]
Our next step is a $\myrho''$--name $q^{(4\omega)}$ for
$x=\tfrac{1}{4}$ giving rise to $p^{(4\omega)}$ with
infinitely many columns containing alternating
quintuples. This is obtained by repeating the arguments
in iv) to obtain initial segments of (variants of) $q^{(3\omega)}$,
stacking them horizontally---in order to obtain an infinity
of columns with alternating quadruples---while extending in top direction
with $\tfrac{1}{4}$; see the middle part of Figure~\ref{figOmega3}.
This forces $\calM$ to output a $\myrho''$--name
$q^{(4\omega)}$ for $\barH(\tfrac{1}{4})=0$ and thus
with in almost every column almost every entry being $\leq\tfrac{1}{4}$,
thus extending the alternating quadruples to quintuples.
\item[vi)]
Noticing that the vertical extension in v) was similar to step
iii), we now take a step similar to iv) based on horizontally
stacked initial parts
of scaled counterparts of $q^{(4\omega)}$ in order to obtain
a $\myrho''$--name $q^{(5\omega)}$ for $x=0$ which $\calM$ maps to
some $p^{(5\omega)}$ containing infinitely many alternating six-tuples.

Then again construct a $\myrho''$--name $q^{(6\omega)}$ for
$x=\tfrac{1}{8}$ by horizontally stacking initial segments of
(variants of) $q^{(5\omega)}$ while extending them vertically
with $\tfrac{1}{8}$ and so on.
\end{enumerate}
Now for the bottom line:
By proceeding the above construction, one eventually obtains
a rational double sequence $q^{(\omega^2)}$ with
$\lim_j q^{(\omega^2)}_{i,j}=0$ for all $i$ ---
that is, a $\myrho''$--name for $x=0$ ---
mapped by $\calM$
to some $p^{(\omega^2)}$ containing (infinitely many) columns $\#k$
with infinitely many alternating entries
$\leq\tfrac{1}{3}$ and $\geq\tfrac{2}{3}$ ---
contradicting that, for $\myrho''$-names $p=(p_{k,\ell})$,
$\lim_\ell p_{k,\ell}$ is required to exist for every $k$.
\qed
\end{proof}
\section{Hierarchies of Hypercomputable Real Functions}
\label{secHierarchies}
The present section investigates and compares the first levels
of the two hierarchies of hypercomputable real functions induced
by the two approaches to real function hypercomputation considered
in Section~\ref{secContinuous}:
based on oracle support and based on weakened encodings.
\subsection{Weakly Computable Real Functions} \label{secWeakly}
For every $(\alpha\to\beta)$--computable function $f:A\to B$,
one may obviously replace representation $\alpha$ for $A$
by a stronger one and $\beta$ for $B$ by a weaker one
while maintaining computability of $f$:
\\[0.5ex]
$\displaystyle
f \quad (\alpha\to\beta)\text{--computable}
  \;\wedge\;\;\alpha,\reduceq\alpha
  \;\;\wedge\;\beta\reduceq\beta' \quad\Rightarrow\quad
    f \quad (\alpha,\to\beta')\text{--computable}$.
\\[0.5ex]
However if both $\alpha$ and $\beta$ are made, say,
weaker then $(\alpha'\to\beta')$--computability
of $f$ may in general be violated.
For $\alpha=\beta=\myrhol$, though, we have seen
in Example~\ref{exHeaviside} that the implication
``$(\myrhol\to\myrhol)\Rightarrow(\myrhol'\to\myrhol')$''
does hold at least for the case of $f$ being
Heaviside's function. By the following result,
it holds in fact for every $f$:
\begin{theorem}[Second Main Theorem of Real Hypercomputation]
\label{thSufficient}
~ \\ Consider $f:\IR\to\IR$.
\begin{enumerate}
\item[a)]
 If $f$ is $(\myrho\to\myrho)$--computable,
 then it is also $(\myrho'\to\myrho')$--computable.
\item[b)]
 If $f$ is $(\myrho\to\myrhol)$--computable,
 then it is also $(\myrho'\to\myrhol')$--computable.
\item[c)]
 If $f$ is $(\myrhol\to\myrhol)$--computable,
 then it is also $(\myrhol'\to\myrhol')$--computable.
\item[d)]
 If $f$ is $(\myrho'\to\myrho')$--computable,
 then it is also $(\myrho''\to\myrho'')$--computable.%
\item[e)]
 If $f$ is $(\myrho''\to\myrho'')$--computable,
 then it is also $(\myrho'''\to\myrho''')$--computable.
\end{enumerate}
The claims remain valid under oracle-supported computation.
\end{theorem}

\medskip
\noindent
As a consequence, we obtain
the following partial strengthening of Lemma~\ref{lemRepr}:
\begin{corollary} \label{corRepr}
~ It holds ~
 $\displaystyle\myrho\;\:\equiv\;\:\,
 \myrhol\!\myjoin\myrhog\;\:\reducneqt\;\:
 \myrhol\;\:\reducneqt\;\myrho'\;\:\equiv\;\:\,
 \myrhol'\!\myjoin\myrhog'\;\:\reducneqt\;\:
 \myrhol'\;\:\reducneqt\;\:\myrho''$ \\
where ``\,$\reduceqt\!$'' denotes \emph{continuous} reducibility
of representations {\rm\mycite{Def.~2.3.2}{Weihrauch}}.
\end{corollary}
\begin{proof}
The positive claims follow from Lemmas~\ref{lemRepr}
and \ref{lemRelative}.
For a negative claim like ``$\myrhol'\notreduceqt\myrho'$''
suppose the contrary. Then by Lemma~\ref{lemRelative},
with the help of some appropriate oracle $\calO$,
one can convert $\myrhol'$--names to $\myrho'$--names.
As Heaviside's function $\Heavi$ is $(\myrho'\to\myrhol')$--computable
by Example~\ref{exHeaviside} and Theorem~\ref{thSufficient},
composition with the presumed conversion
implies $(\myrho'\to\myrho')$--computability of
$\Heavi$ relative to $\calO$---contradicting Theorem~\ref{thNecessary}c).
\qed\end{proof}

\begin{proof}[Theorem~\ref{thSufficient}d]
Let $f$ be $(\myrho'\to\myrho')$--computable and
$x$ given by a $\myrho''$--name, that is, a rational
sequence $q=(q_n)$ with
$x=\lim_i \lim_j q_{\scriptscriptstyle\langle i,j\rangle}$.
For each $i$, compute by assumption from the $\myrho'$--name
$q_{\scriptscriptstyle\langle i,\cdot\rangle}=
(q_{\scriptscriptstyle\langle i,j\rangle})_{_j}$ of
$x_i:=\lim_j q_{\scriptscriptstyle \langle i,j\rangle}$
a $\myrho'$--name of $f(x_i)$, that is, a sequence
$p=p_{\scriptscriptstyle\langle i,\cdot\rangle}=
(p_{\scriptscriptstyle \langle i,j\rangle})_{_j}$ with
$f(x_i)=\lim_j p_{\scriptscriptstyle\langle i,j\rangle}$.
Continuity of $f$ due to Fact~\ref{facFolklore}c) asserts
$$\lim_i \lim_j p_{\scriptscriptstyle\langle i,j\rangle}
\;=\;\lim_i f(x_i)\;\overset{!}{=}\;f\big(\lim_i x_i\big)
\;=\;f\big(\lim_i \lim_j
q_{\scriptscriptstyle\langle i,j\rangle}\big)\;=\;f(x) $$
this sequence $p$ to be a $\myrho''$--name
for $y=f(x)$.
\qed\end{proof}

Where the last proof exploited Fact~\ref{facFolklore}c),
the next one relies on Theorem~\ref{thNecessary}c):

\begin{proof}[Theorem~\ref{thSufficient}e]
A $\myrho'''$--name for $x\in\IR$ is a rational sequence $a=(q_n)$ with
$x=\lim_i\lim_j\lim_k q_{\scriptscriptstyle\langle i,j,k\rangle}$.
For each $i$, exploit $(\myrho''\to\myrho'')$--computability of $f$
to obtain, from the $\myrho''$--name
$q_{\scriptscriptstyle\langle i,\,\cdot\,,\,\cdot\,\rangle}$ of
$x_i:=\lim_j\lim_k q_{\scriptscriptstyle \langle i,j,k\rangle}\in\IR$,
a sequence
$p_{\scriptscriptstyle\langle i,\,\cdot\,,\,\cdot\,\rangle}$
with $\lim_j\lim_k p_{\scriptscriptstyle\langle i,j,k\rangle}$
as $\myrho''$--name of $f(x_i)$.
Similarly to case d), this sequence $p$ constitutes a 
$\myrho'''$--name for $y=f(x)$
by continuity of $f$ due to Theorem~\ref{thNecessary}c).
\qed\end{proof}

\begin{proof}[Theorem~\ref{thSufficient}a]
Let $f$ be $(\myrho\to\myrho)$--computable. Its
$(\myrho'\to\myrho')$--computability is established as follows:
Given $(q_n)\subseteq\IQ$ with $x=\lim_n q_n$,
apply the assumption to evaluate
$f(q_n)$ for each $n$ up to error $2^{-n}$; that is,
obtain $p_n\in\IQ$ with $|p_n-f(q_n)|\leq2^{-n}$.
Since $f$ is continuous by
Fact~\ref{facFolklore}a), it follows
$f(x)=\lim_n f(q_n)=\lim_n p_n$
so that $(p_n)$ is a $\myrho'$--name for $y=f(x)$.
\qed\end{proof}

\noindent
It is interesting 
that the latter proof works in fact uniformly in $f$,
i.e., we have%
\begin{scholium}
The \emph{apply} operator
~$C(\IR)\times\IR\ni(f,x)\mapsto f(x)$~
is $([\myrho\myto\myrho]\times\myrho'\to\myrho')$--computable.
\end{scholium}
Similarly, Theorem~\ref{thSufficient}b) follows from
Lemma~\ref{lemLowersemi} below together with the observation
that every $(\myrho\to\myrhol)$--computable $f$ has a
computable $[\myrho\myto\myrhol]$--name
\cite[\textsc{Corollary~5.1(2)} and \textsc{Theorem~3.7}]{SemiTCS}; 
here, $[\myrho\myto\myrhol]$ denotes a natural representation
for the space $\LSC(\IR)$ of lower semi-continuous
functions $f:\IR\to\IR$ considered in \cite{SemiTCS}.
Specifically, a $[\myrho\myto\myrhol]$--name for such a $f$
is an enumeration of all rational triples $(a,b,c)$ such that
$c<\min f\big[[a,b]\big]$---the latter making sense as a
lower semi-continuous function attains its minimum (though
not necessarily its maximum) on any compact set.
$[\myrho\myto\myrhol]$ indeed \emph{is} a representation for
$\LSC(\IR)$ because different lower semi-continuous functions
give rise to different such collections $\{(a,b,c)\in\IQ^3:\ldots\}$;
cf.~\mycite{Lemma~3.3}{SemiTCS}.
\begin{lemma} \label{lemLowersemi}
~ $\LSC(\IR)\times\IR\ni(f,x)\mapsto f(x)$~
is ~$([\myrho\myto\myrhol]\times\myrho'\to\myrhol')$--computable.
\end{lemma}
\begin{proof}
Let $(a_k,b_k,c_k)_{_k}$ denote the given $[\myrho\myto\myrhol]$--name
of $f\in\LSC(\IR)$ and $(q_n)_{_n}$ the given $\myrho'$--name for $x\in\IR$.
Our goal is to $\myrhol'$--compute $y:=f(x)$.
Define the sequence $p=(p_m)_{_m}\subseteq\IQ\cup\{+\infty\}$ by
\begin{equation} \label{eqLowersemi}
p_{_{\langle k,\ell,n\rangle}}
\;:=\; \left\{ \begin{array}{ll}
\max\big\{c_m \;: \;m\leq k \;\wedge& [a_m,b_m]\supseteq[a_k,b_k]\big\}  \\[0.2ex]
   &\text{ if } q_n\in(a_k,b_k) \;\wedge\; |b_k-a_k|=2^{-\ell} \\[0.5ex]
+\infty &\text{ otherwise} \end{array} \right.
\end{equation}
From the given information, one can
obviously compute $p$. Moreover this sequence satisfies
\begin{itemize}
\item[\labelitemii] ~$\liminf p\geq y$: \\
  Let $\epsilon>0$ be arbitrary. Since $f$ is lower semi-continuous,
  its preimage $f^{-1}\big[(y-\epsilon,\infty)\big]\ni x$ is an open set
  and therefore contains an entire ball around $x$. In fact,
  the center of this ball may be chosen as rational
  and its diameter of the form $2^{-L}$ for some $L\in\IN$;
  formally (see Figure~\ref{figUnifival}): 
  \begin{multline} \label{eqUnifival}
  \exists K,L,K'\in\IN: \quad
    x\in(a_{K'},b_{K'})\subseteq[a_K,b_K]\subseteq
      f^{-1}\big[(y-\epsilon,\infty)\big] \;\;\wedge\\
   |b_K-a_K|=2^{-L} \;
      \wedge\; a_{K'}=a_K+\tfrac{3}{2}\cdot 2^{-L-2}
      \;\wedge\; b_{K'}=b_K-\tfrac{3}{2}\cdot 2^{-L-2}
  \end{multline}
  where we have exploited that \emph{every} rational pair
  $(a,b)$ occurs in the list representing the $[\myrho\myto\myrhol]$--name.
  Moreover, as it consists of all rational triples $(a,b,c)$
  with $c<\min f\big[[a,b]\big]$,
  \begin{equation} \label{eqLowersemi2}
  \exists M\geq K: \;\; [a_K,b_K]=[a_M,b_M] \;\wedge\;
    c_M\geq\min f\big[[a_M,b_M]\big]-\epsilon
    \overset{\text{(*)}}{\;\geq\;}y-2\epsilon
  \end{equation}
  with (*) a consequence of
  $[a_K,b_K]\subseteq f^{-1}\big[(y-\epsilon,\infty)\big]$
  in Equation~(\ref{eqUnifival}). Finally,
  \begin{equation} \label{eqLowersemi3}
  \lim_n q_n=x\in(a_{K'},b_{K'}) 
  \qquad\Rightarrow\quad
  \exists N: \;\forall n\geq N: \;\; q_n\in(a_{K'},b_{K'})\enspace .
  \end{equation}
  So putting things together, for each $n\geq N$, $\ell\geq L'$, and $k\geq M$,
  we either have $p_{_{\langle k,\ell,n\rangle}}=+\infty\geq y-2\epsilon$;
  or we are in the first case of Equation~(\ref{eqLowersemi}), thus%
\begin{itemize}
\item[\labelitemi] $q_n\in(a_k,b_k)$ with $|b_k-a_k|\leq2^{-\ell}$
\item[\labelitemi] $q_n\in(a_{K'},b_{K'})$ by Equation~(\ref{eqLowersemi3})
\item[\labelitemi] hence $[a_k,b_k]\subseteq[a_K,b_K]$
  by Equation~(\ref{eqUnifival}) due to $\ell\geq L'$;
  cf. Figure~\ref{figUnifival}.
\item[\labelitemi] So $[a_k,b_k]\subseteq[a_M,b_M]$ by Equation~(\ref{eqLowersemi2})
\item[\labelitemi] implying $p_{_{\langle k,\ell,n\rangle}}\geq c_M\geq y-2\epsilon$
  by Equations~(\ref{eqLowersemi}) and (\ref{eqLowersemi2}) since $k\geq M$.
\end{itemize}
  Summarizing, it holds $p_{_{\langle k,\ell,n\rangle}}\geq y-2\epsilon$
  for all $(k,\ell,n)\in\IN^3$ not belonging to the finite set
  $\{0,1,\ldots,N-1\}\times\{0,1,\ldots,L'-1\}\times\{0,1,\ldots,M-1\}$
  of exceptions. Consequently $\liminf p\geq y-2\epsilon$;
  even $\liminf p\geq y$ because $\epsilon>0$ was arbitrary.
\begin{figure}[hbt]
\centerline{\includegraphics[width=0.8\textwidth]{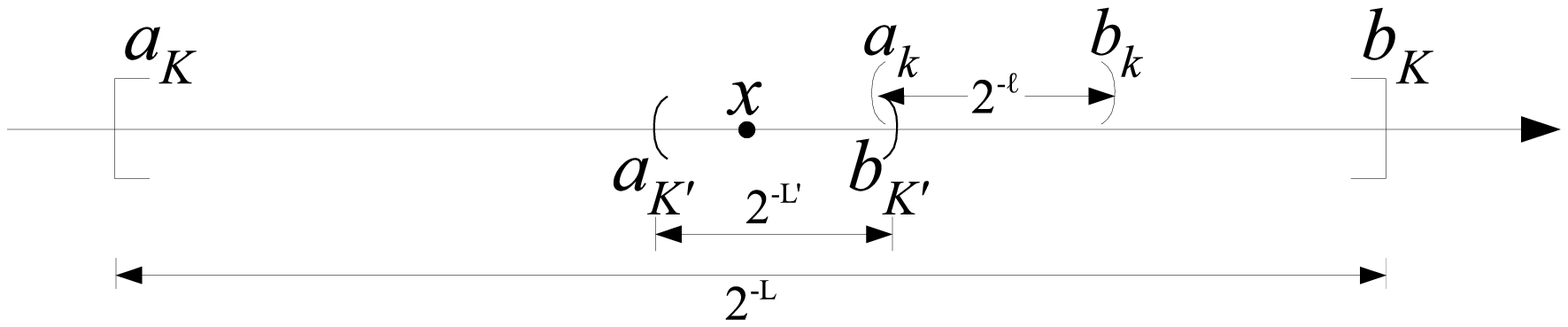}\vspace*{-3ex}}%
\caption{\label{figUnifival}Nesting of some rational intervals of
dyadic length contained in $f^{-1}\big[(y-\epsilon,\infty)\big]$. \newline
The parameters are chosen in such a way that,
whenever $(a_{K'},b_{K'})$ meets some other
$(a_k,b_k)$ of length $|b_k-a_k|=2^{-\ell}$
for $\ell\geq L':=L+2$, then $[a_k,b_k]$ is entirely contained
within the larger $[a_K,b_K]$.}%
\end{figure}
\item[\labelitemii] ~$\liminf p\leq y$ \\
  Indeed: Since the $[\myrho\myto\myrhol]$--name contains in particular all
  rational pairs $(a_k,b_k)$ and these intervals are dense in $\IR$,
  there exists to every $\ell\in\IN$ some $k$ such that
  $|b_k-a_k|=2^{-\ell}$ and $x\in(a_k,b_k)$. Furthermore
  it holds $q_n\in(a_k,b_k)$ for some sufficiently large $n$
  because $\lim_n q_n=x$.
  We have thus infinitely many triples $(k,\ell,n)$ for which
  $p_{_{\langle k,\ell,n\rangle}}$ is defined by the first
  case in Equation~(\ref{eqLowersemi}) and thus agrees with
  some $c_m<\min f\big[[a_m,b_m]\big]\leq f(x)=y$ as
  $x\in(a_k,b_k)\subseteq[a_m,b_m]$.
\end{itemize}
Concluding, we have $\liminf_m p_m=y$.
Although $p$ may attain the value $+\infty$, this can easily
be overcome by proceeding to
$\tilde p_m:=p_m$ for $p_m\not=\infty$
and $\tilde p_m:=\max\{0,\tilde p_0,\ldots,\tilde p_{m-1}\}$ for
$p_m=\infty$ because this transformation $p\mapsto\tilde p$
on sequences obviously does not affect their $\liminf<\infty$.
This yields a $\tildemyrhol'$--name for $y$
which can finally be converted to
the desired $\myrhol'$--name due to the easy part of
Scholium~\ref{schLiminf}.
\qed
\end{proof}

In order to obtain a similar uniform claim yielding
Theorem~\ref{thSufficient}c), recall that every 
$(\myrhol\to\myrhol)$--computable function $f:\IR\to\IR$
is necessarily both monotonically increasing and
lower semi-continuous
(Fact~\ref{facFolklore}b+c). This suggests
\begin{definition} \label{defMLSC}
Let $\MLSC{\IR}$ denote the class of all
monotonically increasing, lower semi-continuous functions
$f:\IR\to\IR$. A $[\myrhol\myto\myrhol]$--name
for $f\in\MLSC{\IR}$ is an enumeration of the set
$\{(a,c)\in\IQ^2: c<f(a)\}$.
\end{definition}
\begin{lemma} \label{lemMLSC}
\begin{enumerate}
\item[a)] Distinct $f,g\in\MLSC{\IR}$ have
  different sets $\{(a,c):\ldots\}$ according to
  Definition~\ref{defMLSC}; that is,
  $[\myrhol\myto\myrhol]$ constitutes a well-defined
  representation.
\item[b)] A function $f\in\MLSC{\IR}$ is
  $(\myrhol\to\myrhol)$--computable iff it
  has a computable $[\myrhol\myto\myrhol]$--name.
\item[c)] Let $f\in\MLSC{\IR}$, $(a_k,c_k)_{_k}$ with
  $\{(a,c)\in\IQ^2:c<f(a)\}=\{(a_k,c_k):k\in\IN\}$,
  $x\in\IR$, and $q=(q_n)\subseteq\IQ$ with
  $x=\liminf_n q_n$. Then, the rational sequence $p$ defined by
  $$p_{_{\langle k,n,\ell\rangle}}
    \;:=\;\left\{ \begin{array}{ll}
\max\big\{c_m \;: \;m\leq k \;\wedge\; a_m\geq a_k\big\}   
  &\text{ ~if~ } a_k<q_n<a_k+2^{-\ell} \\[0.5ex]
+\infty &\text{ ~otherwise} \end{array} \right. $$
  satisfies ~$\liminf p=f(x)=:y$.
\item[d)] Therefore, the \emph{apply} operator
  ~$\MLSC{\IR}\times\IR \;\ni\; (f,x)
  \;\mapsto\; f(x)$~ is
  ~$([\myrhol\myto\myrhol]\times\myrhol'\to\myrhol')$--computable.
\end{enumerate}
\end{lemma}
\begin{proof}
\begin{enumerate}
\item[a)]
Let $f,g\in\MLSC{\IR}$ with $f\not=g$,
that is, w.l.o.g. $f(x_0)<g(x_0)$ for some $x_0\in\IR$.
There exists some $c_0\in\IQ$ with $f(x_0)<c_0<g(x_0)$.
Being monotonically increasing and lower semi-continuous,
their pre-images $f^{-1}\big[(c_0,\infty)\big]\not\ni x_0$
and $g^{-1}\big[(c_0,\infty)\big]\ni x_0$ on open half-interval
$(c_0,\infty)$ are again open half-intervals
$(x_f,\infty)$ and $(x_g,\infty)$, respectively.
As $x_0$ belongs to the second but not to the first,
we have $x_g<x_0<x_f$ and therefore
$x_g<a_0<x_f$ for some $a_0\in\IQ$.
Then $a_0\in (x_g,\infty)=g^{-1}\big[(c_0,\infty)\big]$
yields $c_0<g(a_0)$
whereas $a_0\not\in (x_f,\infty)=f^{-1}\big[(c_0,\infty)\big]$
asserts $c_0\not< f(a_0)$.
\item[b)]
Let $\calM$ denote a Type-2 Machine $(\myrhol\to\myrhol)$--computing
$f\in\MLSC{\IR}$. Evaluating $f$ at $a\in\IQ$ by simulating $\calM$
on the $\myrhol$--name $(a,a,a,\ldots)$ for $a$ thus yields
a $\myrhol$--name for $f(a)$ which is (equivalent to) a list
of all $c\in\IQ$ with $c<f(a)$ \mycite{Lemma~4.1.8}{Weihrauch}. So
dove-tailing this simulation for all $a\in\IQ$ yields the
desired $[\myrhol\myto\myrhol]$--name for $f$.

Conversely, knowing a $[\myrhol\myto\myrhol]$--name
$(a_k,c_k)_{_k}$ for $f\in\MLSC{\IR}$ and given an
increasing sequence $(q_n)\subseteq\IQ$ with $x=\sup_n q_n$,
let
$$p_n:=c_n \text{ ~if~ } a_n\leq q_n,
\qquad p_n:=-\infty \text{ ~otherwise} \enspace . $$
Then, in the first case, $p_n=c_n<f(a_n)\leq f(q_n)\leq f(x)=:y$
by monotonicity, and $p_n=-\infty\leq y$ in the second;
hence $\sup_n p_n\leq y$. To see $\sup_n p_n\geq y$,
fix arbitrary $\epsilon>0$ and consider
the open half-interval
$f^{-1}\big[(y-\epsilon,\infty)\big]=(x_\epsilon,\infty)$
containing $x$ and thus also some rational $a=a_K\in(x_\epsilon,x)$,
$K\in\IN$. Furthermore $q_n\nearrow x$ yields some $N\in\IN$
such that $q_n\in(a_K,x)$ for all $n\geq N$.
And finally there exists $M\geq N$ with
$a_M=a_K$ and $c_M\geq f(a_M)-\epsilon$.
Together this asserts $q_M>a_K=a_M$ because $M\geq N$
and thus $p_M=c_M\geq f(a_K)-\epsilon>y-2\epsilon$
due to $a_k\in f^{-1}\big[(y-\epsilon,\infty)\big]$.
\item[c)]
Take arbitrary $\epsilon>0$. As $f$ is increasing and
lower semi-continuous, the pre-image $f^{-1}\big[(y-\epsilon,\infty)\big]$
is an open half-interval $(x_\epsilon,\infty)$ containing $x$.
Therefore there exist $K,L\in\IN$ such that
$x_\epsilon<a_K$ and $a_K+2^{-L}<x$;
furthermore, the sequence $(a_k,c_k)_{_k}$ containing
\emph{all} rational pairs $(a,c)$ with $c<f(a)$,
there is $M\geq K$ such that $a_M=a_K$ and
$c_M\geq f(a_M)-\epsilon$;
and finally, since $\liminf_n q_n=x>a_M+2^{-L}$,
it holds $q_n>a_M+2^{-L}$ for all $n\geq N$ with an appropriate $N\in\IN$.
Observe that $q>a_M+2^{-L}$ and $a<q<a+2^{-L}$ implies
$a\geq a_M$; so
together we have for all $n\geq N$, $\ell\geq L$, and $k\geq M$
that $p_{_{\langle k,n,\ell\rangle}}$ is either $+\infty$
or $\geq c_M\geq f(a_K)-\epsilon
\geq f(x_\epsilon)-\epsilon\geq y-2\epsilon$
due to monotonicity of $f$ and by definition of $x_\epsilon<a_K$.
This proves $\liminf p\geq y$ because $\epsilon$ was arbitrary.

To see the reverse inequality
``$\liminf p\leq y$'', take arbitrary $\ell\in\IN$.
There exists $k\in\IN$ with $a_k<x<a_k+2^{-\ell}$
and, because of $\liminf_n q_n=x$,
also $n\in\IN$ with $a_k<q_n<a_k+2^{-\ell}$.
We therefore have infinitely many triples $(n,k,\ell)$
for which $p_{_{\langle n,k,\ell\rangle}}$ agrees with
a certain $c_m<f(a_m)\leq f(a_k)\leq f(x)=y$.
\item[d)]
Given a $\myrhol'$--name for $x$, one can obtain a
sequence $(q_n)\subseteq\IQ$ with $x=\liminf_n q_n$
by virtue of Scholium~\ref{schLiminf}.
From this, the sequence $p\subseteq\IQ$ with
$\liminf p=f(x)$ according to c) is obviously
computable and yields, again by Scholium~\ref{schLiminf},
a $\myrhol'$--name for $y=f(x)$.
\qed\end{enumerate}\end{proof}
Concluding this subsection, the classes of
$(\myrho^{(d)}\to\myrho^{(d)})$--computable
real functions $f:\IR\to\IR$ form,
for $d=0,1,\ldots$ respectively,
a hierarchy. By Fact~\ref{facRAH},
this hierarchy is strict as can be seen from
the constant functions $f(x)\equiv c$
with $c\in\RDelta{d+1}$.
\subsection{Arithmetic Weierstra\ss{} Hierarchy} \label{secWeier}
Section~\ref{secWeakly} established the
sequence ~$\myrho, \myrho', \myrho'', \ldots$~
of increasingly weaker representations for $\IR$
to yield the strict hierarchy of
$(\myrho\to\myrho)$--computable,
$(\myrho'\to\myrho')$--computable,
and $(\myrho''\to\myrho'')$--computable
functions $f:[0,1]\to\IR$.
We now compare these classes with those induced
by the other kind of real hypercomputation suggested
in Section~\ref{secContinuous}: relative to
the Halting Problem $H=\emptyset'$ and its
iterated jumps $\emptyset''$, \ldots 

Such a comparison makes sense because both
weakly and oracle-computable real functions
are necessarily continuous according to
Fact~\ref{facFolklore}d)/Theorem~\ref{thNecessary}c)
and Lemma~\ref{lemRelative}, respectively.

\medskip
The classical \person{Weierstra\ss} Approximation Theorem
establishes any continuous real function $f:[0,1]\to\IR$
to be the uniform limit $f=\ulim_n P_n$
of a sequence of rational polynomials $(P_n)\subseteq\IQ[X]$.
Here, `$\ulim$' suggestively denotes uniform convergence of continuous
functions on $[0,1]$, that is the requirement
$$\sup\limits_{0\leq x\leq1}|f(x)-P_n(x)| \;\;=:\;\; \|f-P_n\|\;\to\;0
\qquad\text{as~ } n\to\infty \enspace . $$
The famous \aname{Effective Weierstra\ss{} Theorem}
due to \person{Pour-El}, \person{Caldwell}, and \person{Hauck}
relates effectively evaluable to effectively approximable
real functions:
\begin{fact} \label{facWeier}
A function $f:[0,1]\to\IR$ is $(\myrho\to\myrho)$--computable
if and only if it holds%
\\[0.5ex]\begin{tabular}{rl}
~$\pmb{[\myrho\myto\myrho]}$\textbf{:} &
There exists a computable sequence of
~(degrees and coefficients of) \\
& rational polynomials $(P_n)\subseteq\IQ[X]$ such that%
\hfill%
\makebox[9.0em][l]{\rule[1ex]{0pt}{1.3ex}\smash{\begin{minipage}[b]{10.3em}
\begin{equation} \label{eqWeier}
\|f-P_n\| \;\leq\; 2^{-n}
\end{equation}
\end{minipage}}}%
\end{tabular}%
\end{fact}%
\begin{proof}
See \mycite{Section~0.7}{PER}, \cite{WeierA}, or \cite{WeierB}. \\
The notion ``$[\myrho\myto\myrho]$'' is justified as the list $(P_n)_{_n}$
constitutes (or is equivalent to) a $[\myrho\myto\myrho]$--name for 
$f=\ulim_n P_n$; cf. \cite[bottom of p.160]{Weihrauch}.%
\qed\end{proof}
The aforementioned other approach to continuous real hypercomputation
arises from allowing the fast convergent sequence $(P_n)_{_n}\subseteq\IQ[X]$
to be computable in $\emptyset'$ or $\emptyset''$.
The $\emptyset'$--computable $f:[0,1]\to\IR$ have in fact
already been characterized by \person{Ho} as Claim~a) of the following
\begin{lemma} \label{lemWeier} 
\begin{enumerate}
\item[a)]
To a real function $f:[0,1]\to\IR$, there exists a $\emptyset'$--computable
sequence of polynomials $(P_n)$ satisfying Equation~(\ref{eqWeier})
~if and only if~ it holds
\\[0.5ex]\begin{tabular}{rl}
$\pmb{[\myrho\myto\myrho]'}$\textbf{:} &
There is a computable sequence $(Q_m)\subseteq\IQ[X]$
converging uniformly \\ & (although not necessarily `fast')
to $f$, that is, with $f=\ulim\limits_{m\to\infty} Q_m$.
\end{tabular}
\item[b)]
For an arbitrary oracle $A$, the sequence
(of discrete degrees and numerators/denominators of the coefficients of)
~$(P_n)_{_n}\subseteq\IQ[X]$~
is $A'$--computable ~iff~
there exists an $A$--computable sequence ~$(Q_{n,m})_{_{n,m}}\subseteq\IQ[X]$~
such that
$$ \forall n \;\exists M\;\forall m\geq M: \quad P_n=Q_{n,m} \enspace . $$
\item[c)]
To a real function $f:[0,1]\to\IR$, there exists a $\emptyset''$--computable
sequence of polynomials $(P_n)$ satisfying Equation~(\ref{eqWeier})
~if and only if~ it holds
\\[0.5ex]\begin{tabular}{rl}
$\pmb{[\myrho\myto\myrho]''}$\textbf{:} &
There is a computable sequence $(Q_m)\subseteq\IQ[X]$
s.t.
$f=\ulim\limits_i \ulim\limits_j Q_{\scriptscriptstyle\langle i,j\rangle}$.%
\end{tabular}
\end{enumerate}
\end{lemma}
Notice the similarity of Claims~a+c) to Fact~\ref{facRAH}b).
\begin{proof}
\begin{enumerate}
\item[a)] See \mycite{Theorem~16}{Ho}.
\item[b)] is a straight-forward extension of
  \person{Shoenfield}'s Limit Lemma \mycite{Lemma III.3.3}{Soare}
  and its generalization to sequences of rational numbers
  \mycite{Lemma 4.1}{Zheng}.
\item[c)] 
  If $\emptyset''$--computable ~$(P_n)_{_n}\subseteq\IQ[X]$~ 
  satisfies Equation~(\ref{eqWeier}), then by virtue of
  the relativization of \mycite{Theorem~16}{Ho} there exists some
  $\emptyset'$--computable ~$(\tilde P_n)_{_n}\subseteq\IQ[X]$~
  converging to the same $f$ uniformly on $[0,1]$.
  By Claim~a) in turn, $\tilde P_n=\ulim_m Q_{n,m}$ for some
  computable sequence $(Q_{n,m})\subseteq\IQ[X]$.

  Conversely if $f=\ulim_n\tilde P_n$ with $\tilde P_n:=\ulim_m Q_{n,m}$ 
  for a computable $(Q_{n,m})$, then let $P_n:=Q_{n,m_n}$ where
  \begin{equation} \label{eqFastsub}
  m_n\quad:=\quad \min\big\{m\in\IN \:\big|\,
    \forall k,\ell\geq m \,:\, \|Q_{n,k}-Q_{n,\ell}\|\leq2^{-n}\big\} 
  \enspace .
  \end{equation}
  This sequence $(m_n)_{_n}$ is well-defined and yields
  ~$\|P_n-\tilde P_n\|\leq2^{-n}$, so $f=\ulim_n\tilde P_n=\ulim_n P_n$.
  Moreover, the minimum in Equation~(\ref{eqFastsub}) is taken
  over a co-r.e. set ---
  $r:=\|Q_{n,k}-Q_{n,\ell}\|\cdot2^n$~ being ~$\myrho$--computable
  by virtue of \mycite{Corollary~6.2.5}{Weihrauch}
  and the complementary condition ~``$r>1$''~ 
  $\myrho$--r.e. open
  and hence recursive in $\emptyset'$.
  Similar to Equation~(\ref{eqFastsub}), this $\emptyset'$--computable
  sequence $(P_n)_{_n}\subseteq\IQ[X]$
  converging uniformly though just ultimately to $f$
  can be turned into a $\emptyset''$--computable, fast convergent one.
\qed
\end{enumerate}
\end{proof}
We thus have two hierarchies of hypercomputable
continuous real functions:
\\ \noindent \textbullet\quad\,\!
$[\myrho\myto\myrho]$, \;\,\quad $[\myrho\myto\myrho]'$,
  \;\:\,\quad $[\myrho\myto\myrho]''$,
 \quad\ldots
\\ \noindent \textbullet\quad
$(\myrho\to\myrho)$, ~ $(\myrho'\to\myrho')$, ~
  $(\myrho''\to\myrho'')$, ~ \ldots
\\ \noindent
By the Fact~\ref{facWeier},
their respective ground-levels coincide.
Our next result compares their respective higher levels.
They turn out to lie skewly to each other (Claim~c).
\begin{theorem} \label{thWeier}
\begin{enumerate}
\item[a)] Let $f:[0,1]\to\IR$ be $[\myrho\myto\myrho]'$--computable
  (in the sense of Lemma~\ref{lemWeier}a).
  Then, $f$ is $(\myrho'\to\myrho')$--computable.
\item[b)] Let $f:[0,1]\to\IR$ be $(\myrho'\to\myrho')$--computable.
  Then, $f$ is $[\myrho\myto\myrho]''$--computable.
\item[c)] There is a $(\myrho'\to\myrho')$--computable
  but not $[\myrho\myto\myrho]'$--computable $f:[0,1]\to\IR$.
\end{enumerate}
\end{theorem}
The idea to c) is that
every $[\myrho\myto\myrho]'$--computable $f:[0,1]\to\IR$
has a modulus of uniform continuity recursive in $\emptyset'$;
whereas a $(\myrho'\to\myrho')$--computable $f$, 
although uniformly continuous as well, in general does not.

Before proceeding to the proof, 
we first provide a tool which turns out to be useful in the sequel.
It is well-known in Recursive Analysis that, although equality
of real numbers is $\myrho$--undecidable due to the Main Theorem,
inequality is at least semi-decidable. The following lemma
generalizes this to $\myrho'$ and to
$(\myrho'\to\myrhol')$--computable functions:
\begin{lemma} \label{lemBound}
\begin{enumerate}
\item[a)]
  Let $f:\IR\to\IR$ be $(\myrho\to\myrhol)$--computable. Then
  the property
  $$ \big\{ (a,b,c)\in\IQ^3 \,\big|\, \exists x\in[a,b]: f(x)>c \big\} $$
  whether $f$ on $[a,b]$ exceeds $c$ ~ is semi-decidable.
\item[b)]
  Let $f:\IR\to\IR$ be $(\myrho'\to\myrhol')$--computable. Then
  the property
  $$ \big\{ (a,b,c)\in\IQ^3 \,\big|\, \exists x\in[a,b]: f(x)>c \big\} $$
  whether $f$ on $[a,b]$ exceeds $c$ ~ is semi-decidable
  relative to $\emptyset'$.
\item[c)]
  Let $f:\IR\to\IR$ be $(\myrho'\to\myrho')$--computable.
  Then the property
  $$ \big\{ (a,b,c,m)\in\IQ^3\times\IN \,\big|\,
     \forall x\in[a,b]: c-2^{-m}\leq f(x)\leq c+2^{-m}\big\}$$
  is decidable relative to $\emptyset''$.
\end{enumerate}
\end{lemma}
\begin{proof}
a) is standard; c) follows from b) which is established as follows: \\
By lower semi-continuity of $f$ due to Theorem~\ref{thNecessary}a),
if $f$ exceeds $c$ on the compact interval $[a,b]$, then it does so
on some \emph{rational} $x$. Feeding, for any such $x\in[a,b]\cap\IQ$,
the $\myrho'$--name $(x,x,x,x,\ldots)$ for $x$
into the Type-2 Machine computing
$f$ reveals the mapping $\IQ\ni x\mapsto f(x)$ to be
$(\nuQ\to\myrhol')$--computable. With $\emptyset'$--oracle,
it thus becomes $(\nuQ\to\myrhol)$--computable by virtue
of \mycite{Lemma~4.2}{Zheng}. Since
$\{(y,c):y>c\}$ is $(\myrhol\times\nuQ)$--semi-decidable, the claim follows.
\qed
\end{proof}

\begin{proof}[Theorem~\ref{thWeier}]
\begin{enumerate}
\item[a)] Let $(P_n)\subseteq\IQ[X]$ denote a computable sequence
converging uniformly (yet not necessarily fast) to $f$.
Let $x\in[0,1]$ be given as the limit of a sequence $(q_n)\subseteq\IQ$.
Then, $p_n:=P_n(q_n)$ eventually converges to $f(x)$.
\item[b)]
 Let $x\in[0,1]$ be given by (an equivalent to)
 its $\myrho$--name in form of two
 rational sequences $(a_n)$ and $(b_n)$ with
 $\{x\}=\bigcap_n [a_n,b_n]$. There exists a rational
 sequence $(c_m)$ forming a $\myrho$--name for $f(x)$,
 that is, satisfying
 $c-2^{-m}\leq f(x)\leq c+2^{-m}$ for all $m$;
 and by virtue of Lemma~\ref{lemBound}c), such a sequence can
 be found with the help of a $\emptyset''$--oracle.
 This reveals that $f$ is $\emptyset''$--recursive in
 the sense of \mycite{Section~4}{Ho} and thus,
 similarly to \mycite{Corollary~17}{Ho},
 $[\myrho\myto\myrho]''$--computable.
\item[c)] Let $h:\IN\to\IN$ denote a $\emptyset'$--computable
  injective total enumeration of some subset
  $H=h[\IN]\in\Sigma_2\setminus\Delta_2$.
  Observe that $a_m:=2^{-h(m)}$ is a
  $\myrho'$--computable real sequence
  converging to 0 with modulus of convergence
  \mycite{Definition~4.2.2}{Weihrauch} lacking $\emptyset'$--recursivity;
  compare \cite[\textsc{Exercise~4.2.4}c)]{Weihrauch}.
  Let $\varphi:\IR\to\IR$ denote some $(\myrho\to\myrho)$--computable
  unit pulse, that is, vanishing outside $[0,1]$ and having
  height $\max_x\varphi(x)=\varphi(\tfrac{1}{2})=1$; a piecewise linear
  `\emph{hat}' function for instance will do
  fine but we can even choose $\varphi$ as in
  \mycite{Theorem~1.1.1}{PER} to obtain the counter-example
  \begin{equation} \label{eqPulses}
  f(x) \;:=\; \sum_{m\in\IN}  a_m\cdot\varphi(2^m x-1) \enspace ,
  \end{equation}
  (that is, a non-overlapping superposition of scaled translates
  of such pulses) to be $C^\infty$.
  By Theorem~\ref{thSufficient}a),
  $x\mapsto a_m\cdot\varphi(2^m x-1)$ is $(\myrho'\to\myrho')$--computable;
  in fact even uniformly in $m$: Given $(q_n)_{_n}\subseteq\IQ$
  with $x=\lim_n q_n$, one can for each $M\in\IN$ obtain
  a sequence $(p_{k,M})_{_{k}}\subseteq\IQ$ with
  $\lim_k p_{k,M}=\sum_{m\leq\pmb{M}} a_m\cdot\varphi(2^mx-1)=:f_M$.
  The functions $f_M$ converge uniformly (though not effectively) to $f$
  because of the disjoint supports of the terms $\varphi(2^mx-1)$
  in Equation~(\ref{eqPulses}).
  Therefore $\lim\limits_M p_{M,M}=f(x)$, thus establishing
  $(\myrho'\to\myrho')$--computability of $f$.
\\  Suppose $f$ was $[\myrho\myto\myrho]'$--computable.
  Then, by virtue of \mycite{Lemma~15}{Ho}, it has a
  $\emptyset'$--recursive modulus
  of uniform continuity; cf. \mycite{Definition~6.2.6.2}{Weihrauch}.
  In particular given $n\in\IN$, one can $\emptyset'$--compute
  $m\in\IN$ such that
  $x:=2^{-m}$ and $y:=\tfrac{3}{2}x$ satisfy
  \quad$2^{-n} \;{\geq}\;
    |f(x)-f(y)| \;=\; |0-a_m|$\quad
  contradicting that $(a_m)$ has no $\emptyset'$--recursive
  modulus of continuity.
\qed\end{enumerate}
\end{proof}
\section{Type-2 Nondeterminism} \label{secNondet}
Concerning the two kinds of
real hypercomputation considered so far---based on oracle-support
and weak real number encodings that is---recall that the 
according proofs of Fact~\ref{facFolklore} and Theorem~\ref{thNecessary}
crucially rely on the underlying Turing Machines
to behave deterministically.
This raises the question whether nondeterminism
might yield the additional power necessary for evaluating
discontinuous real functions like Heaviside's.

In the discrete (i.e., Type-1) setting where any computation
is required to terminate, the finitely many possible choices
of a nondeterministic machine can of course be simulated by a
deterministic one---however already here subject to the important
condition that all paths of the nondeterministic computation
indeed terminate, cf. \cite{Boas}.
In contrast, a Type-2 computation realizes a transformation
from/to infinite strings and is therefore a generally non-terminating
process. Therefore, nondeterminism here involves an infinite
number of guesses which turns out cannot be simulated by a deterministic
Type-2 machine. 

We also point out that nondeterminism has already before
been revealed not only a useful but indeed the most natural concept
of computation on $\Sigma^\omega$. More precisely,
\person{B\"{u}chi} extended Finite Automata from
finite to infinite strings
and proved that here, as opposed deterministic,
\emph{non}deterministic ones are closed under complement \cite{Buechi}
and thus the appropriate model of computation.
\begin{figure}[htb]
\vspace*{-2.4ex}%
\centerline{\sf\normalsize\begin{tabular}{l||c@{\;}|@{\;}c}
Chomsky-Level & $\displaystyle\Sigma^*$  &  $\displaystyle\Sigma^\omega$ \\[0.2ex] 
\hline
3: regular & Finite Automata &  B\"{u}chi Automata (nondeterministic) \\[0.4ex]
2: context-free & \\[0.4ex]
1: context-sensitive & \\[0.4ex]
0: unrestricted & (Type-1) Turing Machines &  \emph{non}deterministic Type-2 Machines
\end{tabular}}\caption{Models of Computation 
in Chomsky's Hierarchies over
finite/infinite strings\label{figChomsky}}%
\vspace*{-2.2ex}%
\end{figure}
Since automata and Turing Machines
constitute the bottom and top levels, respectively, of \person{Chomsky}'s
Hierarchy of classical languages $L\subseteq\Sigma^*$ ~(Type-1 setting),
we suggest that over infinite strings $\Sigma^\omega$ ~(Type-2 setting)
both their respective counterparts, that is
B\"{u}chi Automata \emph{and Type-2 Machines}
be considered nondeterministically; compare Figure~\ref{figChomsky}.

The concept of nondeterministic computation of a
function ~$f:\subseteq\Sigma^*\to\Sigma^*$~ 
(as opposed to a decision problem) is taken from the famous
\person{Immerman}-\person{Szelepsc\'{e}nyi} Theorem in
computational complexity;
cf. for instance 
\cite[the paragraph preceding \textsc{Theorem~7.6}]{Papadimitriou}:
For $\bar x\in\dom(f)$, some computing paths of
the according machine $\calM$ may fail by leading to rejecting
states, as long as%
\begin{enumerate}
\item[1)] there \emph{is} an accepting computation
of $\calM$ on $\bar x$ \quad and 
\item[2)] \emph{every} accepting computation of
$\calM$ on $\bar x$ yields the correct output ~$f(\bar x)$.
\end{enumerate}
This notion extends straight-forwardly from Type-1 to the Type-2 setting:
\begin{definition} \label{defNondet}
Let $A$ and $B$ be sets with respective representations
$\alpha:\subseteq\Sigma^\omega\to A$ and
$\beta:\subseteq\Sigma^\omega\to B$.
A function $f:\subseteq A\to B$ is called
\emph{nondeterministically} $(\alpha\to\beta)$--computable
if some nondeterministic one-way Turing Machine $\calM$,%
\begin{itemize}
\item upon input of any $\alpha$--name $\bar\sigma\in\Sigma^\omega$
  for some $a\in\dom(f)$,
\item has a computation which outputs a $\beta$--name
  for $b=f(a)$ \hfill\underline{and}
\item every infinite computation\footnote{This condition is slightly
  stronger than the one required in \mycite{Definition~14}{CIE}.}
  of $\calM$ on $\bar\sigma$
  outputs a 
  $\beta$--name for $b=f(a)$.%
\end{itemize}
\end{definition}
This definition is sensible insofar as it leads
to closure under composition:
\begin{observation} \label{obsComposition}
Let $f:\subseteq A\to B$ be nondeterministically 
$(\alpha\to\beta)$--computable
and $g:\subseteq B\to C$ be nondeterministically 
$(\beta\to\gamma)$--computable.
Then, $g\circ f:\subseteq A\to C$ is nondeterministically
$(\alpha\to\gamma)$--computable.
\end{observation}
A subtle point in Definition~\ref{defNondet},
the nondeterministic machine may `withdraw' a 
guess as long as it does so within finite time.
\begin{example}[`Deciding' the Arithmetic Hierarchy] \label{x:Kleene}
Let $P\subseteq\IN$ be recursive,%
\[ A \;=\; \big\{ x\in\IN\,|\, \forall y_1\in\IN \exists z_1\in\IN \;\;\forall y_2 \exists z_2
\ldots \forall y_k \exists z_k : \langle x;y_1,z_1,\ldots,y_k,z_k\rangle\in P\big\} \]
on (or below) level $\Pi_{2k}$ of Kleene's Arithmetic Hierarchy.
Then the function $\tilde\chi_A:\IN\to\{0,1\}\times\{\text{\textvisiblespace}\,\}^{\omega}$ 
is nondeterministically computable:
\\
Observe that $x\in A$ iff
\[ 
\exists f_1,f_2,\ldots,f_k:\IN\to\IN \;
\forall y_1,y_2,\ldots,y_k\in\IN\;:\;
\langle x;y_1,f(y_1),\ldots,y_k,f(y_k)\rangle\in P \]
So given $x\in\IN$, let $\calM_+$ output ``1''
and then verify, while continuously spitting out blanks ``\textvisiblespace'', 
that $\chi_A(x)=1$ indeed holds.
To this end, the machine starts `guessing' the values of
$\bar f=(f_1,\ldots,f_k)$ restricted to $\{0,1,\ldots,n\}$
for $n=1,2,\ldots$
Simultaneously by means of dove-tailing, 
$\calM_+$ tries all $\bar y\in\{0,1,\ldots,n\}^k$
and aborts in case that the assertion
``$\langle x;y_1,f(y_1),\ldots,y_k,f(y_k)\rangle\in P$''
fails.
\\
Now if $x\in A$,
then an appropriate $\bar f$ exists, is ultimately
`found' by $\calM_+$, and leads to indefinite execution; 
whereas if $x\not\in A$, then $\calM_+$ will eventually
terminate for any guessed $\bar f$.
\\
Since $\IN\setminus A\in\Pi_{2k+2}$, a machine
$\calM_-$ can output ``0'' and then similarly 
verify $\chi_A(x)=0$. The final machine $\calM$,
upon input of $x\in\IN$,
nondeterministically chooses to proceed 
either like $\calM_+$ 
or like $\calM_-$. Its computation 
satisfies the requirements
of Definition~\ref{defNondet}.
\qed\end{example}
The power of nondeterministic computation permits 
conversion forth and back among
representations on the Real Arithmetic Hierarchy
from Definition~\ref{defRAH}:
\begin{theorem}[Third Main Theorem of Real Hypercomputation] \label{thNondet}
For each $d=0,1,2,\ldots$, the identity $\IR\ni x\mapsto x$
is nondeterministically $(\myrho^{(d+1)}\to\myrho^{(d)})$--computable.
It is furthermore nondeterministically $(\myrho\to\myrhob)$--computable.
\end{theorem}
\begin{proof}
Consider first the case $d=0$.
Let $x\in\IR$ be given by a sequence $(q_n)\subseteq\IQ$
eventually converging to $x$. Then, there exists a
fast convergent Cauchy sub-sequence $(q_{n_k})_{_k}$,
that is, satisfying
\begin{equation} \label{eqFast}
\forall k\geq\ell:
\quad |q_{n_k}-q_{n_{\ell}}|\;\leq\;2^{-\ell-1}
\end{equation}
and thus forming a $\myrho$--name for $x$.
To find this subsequence, guess iteratively for each $k\in\IN$ 
some $n_k>n_{k-1}$ and check whether it complies with
Inequality~(\ref{eqFast}) for the (finitely many)
$\ell\leq k$; if it does not, we may abort
this computation in accordance with
Definition~\ref{defNondet}.

For $d=1$, let $x=\lim_n x_n$ with $x_n=\lim_m q_{n,m}$.
Then apply the case $d=0$ to
convert for each $n$ the $\myrho'$--name $(q_{n,m})_{_m}$ of $x_n\in\IR$
into an according $\myrho$--name, that is,
a sequence $p_{n,m}$ satisfying $|x_n-p_{n,m}|\leq2^{-m}$.
Its diagonal $(p_{n,n})_{_n}$ then has
$|x-p_{n,n}|\leq|x-x_n|+2^{-n}\to0$ and is thus a $\myrho'$--name
for $x$. Higher levels $d$ can be treated similarly by induction.

For $(\myrho\to\myrhob)$--computability,
let $x\in(0,2)$ be given by a fast convergent sequence $(q_n)\subseteq\IQ$.
We guess the leading digit $b\in\{0,1\}$ for $x$'s binary expansion
$b.*$; in case $b=0$, check whether $x>1$%
---a $\myrho$--semi decidable property---and if so, abort;
similarly in case $b=1$, abort if it turns out that $x<1$.
Otherwise (that is, proceeding while simultaneously continuing the 
above semi-decision process via dove-tailing)
replace $x$ by $2(x-b)$ and repeat guessing the next bit.
\qed\end{proof}
It is also instructive to observe how, in the case of non-unique binary expansion
(i.e., for dyadic $x$), nondeterminism in the above
$(\myrho\to\myrhob)$--computation generates,
in accordance with the third requirement of Definition~\ref{defNondet},
both possible expansions.

Theorem~\ref{thNondet} implies that
\emph{non}deterministic computability of real functions is
largely independent of the representation under consideration
--- in striking contrast to the classical case (Corollary~\ref{corRepr})
where the effectivity subtleties arising from different 
encodings had confused already Turing himself \cite{Turing2}.
\begin{corollary} \label{corIndependent}
\begin{enumerate}
\item[a)] With respect to nondeterministic reduction ~``$\Nreduceq$''~,
it holds ~
 $\displaystyle\myrhob\;\:\Nequiv\;\:\myrho\;\:\Nequiv\;\:
 \myrhol\;\:\Nequiv\;\myrho'\;\:\Nequiv\;\:
 \myrhol'\;\:\Nequiv\;\:\myrho''\;\:\Nequiv\;\ldots$.
\item[b)]
  The entire Real Arithmetic Hierarchy 
  of \person{Weihrauch} and \person{Zheng}
  is nondeterministically computable.
\end{enumerate}
\end{corollary}
\begin{proof}
\begin{enumerate}
\item[a)] follows from Lemma~\ref{lemRepr} and Theorem~\ref{thNondet}.
\item[b)] Let $x\in\RDelta{d+1}$ for some $d\in\IN$.
  Then, $x\in\IR$~ is ~$\myrho^{(d)}$--computable
  by Definition~\ref{defRAH}; hence nondeterministically
  also $\myrhob$--computable by a). Alternatively
  combine Example~\ref{x:Kleene} with Fact~\ref{facRAH}a).
\qed
\end{enumerate}
\end{proof}
In particular, this kind of hypercomputation allows for 
nondeterministic 
$(\myrho\to\myrho)$--evaluation of Heaviside's function
by appending to the $(\myrho\to\myrhol)$--computation in
Example~\ref{exHeaviside} a conversion
from $\myrhol\reduceq\myrho'$ back to $\myrho$.
Section~\ref{secHotz} establishes many more real functions,
both continuous and discontinuous ones,
to be nondeterministically computable, too.
\subsection{Nondeterministic and Analytic Computation} \label{secHotz}
We now show that Type-2 nondeterminism
includes the algebraic so called \aname{BCSS-model} of
real number computation due to \person{Blum}, \person{Cucker}, 
\person{Shub}, and \person{Smale} \cite{BSS,BCSS}
employed for instance in Computational Geometry \mycite{Section~1.4}{CompGeom}.
As a matter of fact, nondeterministic real hypercomputation even covers all
\emph{quasi-strongly $\delta$--$\IQ$--analytic}
functions $f:\subseteq\IR^d\to\IR$ in the sense of
\person{Chadzelek} and \person{Hotz} \mycite{Definition~5}{Hotz}.
The latter can be considered a
synthesis of the Type-2 (i.e., infinite approximate)
and the \BCSS (i.e., finite exact) model of real number computation.
Its computational power admits 
an elegant characterization (see Lemma~\ref{lemHotz}b+c) 
in terms of the following
\begin{definition} \label{defHotz}
A $\myrhoh$--name for $x\in\IR$ is some $(q_n)_{_n}\subseteq\IQ$
such that 
\begin{equation} \label{e:Hotz}
\exists N \;\forall n\geq N: \quad |q_n-x|\leq2^{-n} \enspace .
\end{equation}
\end{definition}
The encoding sequence of rational approximations must thus converge
fast with the exception of some initial segment of finite yet
unknown length. It corresponds to $\myrho$--computation by
an \emph{Inductive} Turing Machine in the sense of
\cite{Burgin} which is roughly speaking a Type-2
Machine but whose output tape(s) need not be one-way
\mycite{Section~2.1}{Weihrauch} provided that
the contents of every cell ultimately stabilizes.
\begin{lemma} \label{lemHotz}
\begin{enumerate}
\item[a)] It holds
~$\myrho\;\reducneq\;\myrhoh\;\reducneq\;\myrho'$.
\item[b)] A function
~$f:\subseteq\IR^N\to\IR$~ is
~$(\myrho^N\to\myrhoh)$--computable
~ iff ~
it is computable by a
quasi-strongly $\delta$--$\IQ$--analytic machine.
\item[c)] $(\myrho\to\myrhoh)$--computability
is equivalent to $(\myrhoh\to\myrhoh)$--computability.
\item[d)] The class of $(\myrhoh\to\myrhoh)$--computable
  functions is closed under composition.
\end{enumerate}
\noindent
The above claims relativize.
\end{lemma}
\begin{proof}
\begin{enumerate}
\item[a)] is immediate. 
\item[b)] Observe that
the \emph{robustness} of the program $\pi$ required 
in \cite[top of p.157]{Hotz} amounts to
the argument $x\in\IR$ of $f$ being accessible
by rational approximations $q_n\in\IQ$ of
error $|q_n-x|\leq2^{-n}$, that is,
in terms of a $\myrho$--name.
The output $y=f(x)$ on the other hand proceeds
by way of two sequences $(p_m)_{_m},(\epsilon_m)_{_m}\subseteq\IQ$
such that $\epsilon_m\to0$ and $|p_m-y|\leq\epsilon_m$
holds for all sufficiently large $m$. 
By effectively proceeding to an appropriate subsequence,
we can w.l.o.g. suppose $\epsilon_m=2^{-m}$,
hence $(p_m)$ is $\myrhoh$--name of $y$.
\item[c)]
By a), every $(\myrhoh\to\myrhoh)$--computable
function is $(\myrho\to\myrhoh)$--computable, too.
For the converse implication, take the Type-2 Machine 
$\calM$
converting $\myrho$--names for $x\in\IR$ to
$\myrhoh$--names for $y=f(x)$. Let $(q_n)$ satisfy
Equation~(\ref{e:Hotz}) for some unknown $N\in\IN$. 

Now simulate $\calM$ on $(q_n)_{_{n\geq0}}$,
implicitly supposing that it is a valid $\myrho$--name,
i.e., that $N=0$. Simultaneously check consistency of
Condition~(\ref{e:Hotz}), that is, 
verify $|q_n-q_k|\leq2^{-n+1}\forall k\geq n\geq N$.
If (or, rather, when) the latter fails for
some $(k_0,n_0)$, $\calM$ has output only finitely 
(say $M_0\in\IN$) many $p_m\in\IQ$. In that case, 
restart $\calM$ on $(q_n)_{_{n\geq1}}$ 
presuming $N=1$ while, again, checking this
presumption consistent with (\ref{e:Hotz});
but this time throw away the first $M_0$
elements of the sequence printed by $\calM$.
Continue analogously for $N=2,3,\ldots$.

We claim that this yields output of a
$\myrhoh$--name for $y$.
Since $(q_n)$ is a valid $\myrhoh$--name, 
a feasible $N$ will eventually be found.
Before that happens, the several partial runs
of $\calM$ have produced only finitely (say $M\in\IN$) 
many rational numbers $p_m$; and after that, 
the final simulation generates
by presumption a valid $\myrhoh$--name for $y$.
Out of this sequence $(p_m)_{_m}$, the first $M$ 
entries may have been exchanged by outputs of 
previous simulation trials; however according to
Definition~\ref{defHotz}, the representation
$\myrhoh$ is immune against such finite 
modifications.
\item[d)]
Quasi-strongly $\delta$--$\IQ$--analytic functions
are closed under composition according to \mycite{Lemma~2}{Hotz};
now apply b+c).
\qed\end{enumerate}\end{proof}
A \BCSS (or, equivalently, an $\IR$--) machine $\calM$
is permitted to store a finite number of \emph{arbitrary}
real constants $r_1,\ldots,r_k$ 
\cite[Instruction 1(b) in \textsc{Table~1} on p.154]{Hotz}
and use it for instance to solve the Halting 
or any other fixed discrete problem \mycite{Example~6}{BSS}.
Slightly correcting \mycite{Theorem~3}{Hotz}, $\calM$'s simulation
by a rational machine thus \emph{requires} knowledge of $\bar r:=(r_1,\ldots,r_k)\in\IR^k$;
e.g. by virtue of oracle access to 
($\calO:=\{\bin(n):\sigma_n=1\}\subseteq\{0,1\}^*$ 
as natural encoding of a $\myrhob^k$--name $\bar\sigma\in\{0,1\}^\omega$ of) 
$\bar r$---compare \cite{Boldi} for the case of
simulating $\myrho$--semi decidability.
\begin{proposition} \label{proHotz}
\begin{enumerate}
\item[a)]
  A function $f:\subseteq\IR\to\IR$ computable
  by a \BCSS--machine with constants $\bar r\in\IR^k$
  is also $(\myrhoh\to\myrhoh)$--computable
  \emph{relative to} $\bar r$.%
\item[b)]
  Every $(\myrho\to\myrho)$--computable function
  $f:\subseteq\IR\to\IR$ 
  is also $(\myrhoh\to\myrhoh)$--computable.
\item[c)]
  Let ~$f:\subseteq\IR\to\IR$~ be $(\myrhoh\to\myrhoh)$--computable
  relative to some oracle $\calO\subseteq\Sigma^*$ in 
  (Kleene's) Arithmetic Hierarchy.
  Then $f$ is nondeterministically Type-2 computable.
\end{enumerate}
\end{proposition}
\begin{proof}
\begin{enumerate}
\item[a)] See (the proof of) \mycite{Theorem~3}{Hotz}.
\item[b)] Combine Lemma~\ref{lemHotz}a+c).
\item[c)] The nondeterministic simulation can answer
  queries to $\calO$ due to Example~\ref{x:Kleene}.
  As $\myrho\Nequiv\myrhoh\Nequiv\myrho'$ by
  Corollary~\ref{corIndependent}a)
and Lemma~\ref{lemHotz}a),
the claim follows.%
\qed\end{enumerate}\end{proof}
Let us illustrate Proposition~\ref{proHotz}a) with the following%
\begin{example} \label{x:Heavi}
Heaviside's Function $h:\IR\to\{0,1\}$ is
trivially \BCSS--computable. 
It is also
$(\myrhoh\to\myrhoh)$--computable by 
means of \emph{conservative branching}: 
Given $x\in\IR$ by virtue of $(q_n)\subseteq\IQ$ with (\ref{e:Hotz})
and unknown $N\in\IN$, let $p_n:=0$ if $q_n\leq2^{-n}$
and $p_n:=1$ otherwise. 
\\
Indeed if $x\leq 0$ then, for all $n\geq N$, 
$q_n\leq2^{-n}$ and thus $p_n=0=f(x)$.
If on the other hand $x>0$, $x>2^{-M}$ for some $M\in\IN$;
then, for all $n\geq\max\{M+1,N\}$,
$q_n>2^{-n}$ so $p_n=1=f(x)$.
\qed\end{example}
%
Of course the class of nondeterministic Type-2 Machines 
(and thus also that of the nondeterministically computable real functions)
is still only countably infinite: most (even constant) functions 
$f:\IR\to\IR$ 
actually remain infeasible to this kind of real hypercomputation.
\section{Conclusion}\label{secConclusion}
Recursive Analysis is often criticized for being unable,
due to its Main Theorem, to
non-trivially treat discontinuous functions.
Although one can in Type-2 Theory devise sensible computability
notions for, say, generalized (and in particular discontinuous)
functions as for instance in \cite{Ning}, evaluation
$x\mapsto f(x)$ of an $L^2$ function or a distribution $f$
at a point $x\in\IR$ does not make sense here
already mathematically.
Regarding the Main Theorem's connection to the Church-Turing Hypothesis
indicated in the introduction, the present work has investigated
whether and which models of hypercomputation allows for
lifting that restriction.

A first idea, relativized computation on oracle Turing Machines,
was ruled out right away. A second idea, computation based on
weakened encodings of real numbers, renders evaluation
~$x\mapsto\Heavi(x)$~ of 
Heaviside's function---although discontinuous---for 
instance $(\myrho\to\myrhol)$--computable.
The drawback of this notion of real hypercomputation:
it lacks closure under composition.
\begin{example} \label{exComposition}
Let $f:\IR\to\IR$, ~ $f(0):=0$ ~and~ $f(x):=1$ for $x\not=0$. 
Let $g(x):=-x$. 
Then both $f$ and $g$ are $(\myrho\to\myrhol)$--computable
but their composition ~ $g\circ f:0\mapsto 0, \;0\not=x\mapsto -1$ ~
lacks lower semi--continuity.
\end{example}
%
Requiring both argument $x$ and value $y=f(x)$ to be encoded
in the same way---say, $\myrho$, $\myrho'$, or $\myrho''$---%
asserts closure under both composition and negation
$f\mapsto -f$; and the prerequisites 
of the Main Theorem applies only to the case $(\myrho\to\myrho)$. Surprisingly, 
$(\myrho'\to\myrho')$--computability and $(\myrho''\to\myrho'')$--computability
still require continuity! 
These results 
extend to $(\myrho^{(d)}\to\myrho^{(d)})$--computability for
arbitrary $d$, although already the step from $d=1$ to $2$ 
made proofs significantly more involved.

These claims immediately relativize, that is, even a mixture of
oracle support and weak real number encodings does not allow
for hypercomputational evaluation of discontinuous functions.
This is due to the purely information-theoretic nature of the 
arguments employed, specifically: the deterministic behavior
of the Turing Machines under consideration.

So we have finally looked at nondeterminism as a further way of
enhancing the underlying machine model beyond Turing's barrier.
Over the Type-2 setting of infinite strings $\Sigma^\omega$, 
this parallels B\"{u}chi's well-established generalization 
of finite automata to so-called \emph{$\omega$--regular} languages.
While the practical realizability of Type-2 nondeterminism
is admittedly even more questionable than that of classical 
$\cal{NP}$-machines, it does yield an elegant notion of
hypercomputation with nice closure properties and
invariant under various encodings.

A precise characterization of the class of nondeterministically
computable real functions will be subject of future work.

\end{document}